\documentclass[journal]{IEEEtran}
\usepackage{cite}      
\usepackage{multirow} 
\usepackage{booktabs}
\usepackage{makecell}
\usepackage{graphicx}  
\usepackage{soul}     
\usepackage{color, xcolor} 
\definecolor{mycolor}{RGB}{7,133,230}  
\definecolor{mycolor1}{RGB}{0,176,80} 
\definecolor{mycolor2}{RGB}{192,0,0}  
\definecolor{mycolor3}{RGB}{45,84,160}  
\usepackage{amsmath}   
\usepackage{amsfonts} 
\usepackage{algorithm}  
\usepackage{algpseudocode} 
\usepackage{cases} 
\usepackage{amssymb} 
\usepackage{amsthm}
\usepackage{bm}
\usepackage{enumitem}
\usepackage{titlesec}
\usepackage{utfsym}
\titlespacing*{\subsubsection}{0pt}{\baselineskip}{0pt}

\usepackage{threeparttable}  
\usepackage{hyperref}  
\hypersetup{ 
	colorlinks=true,
	linkcolor=blue,
	filecolor=blue,      
	urlcolor=blue,
	citecolor=blue,
}
\ifCLASSOPTIONcompsoc
\usepackage[caption=false,font=normalsize,labelfont=sf,textfont=sf]{subfig}
\else
\usepackage[caption=false,font=footnotesize]{subfig}
\fi

\ifCLASSINFOpdf
\else
\fi

\begin{document}
\title{Revitalizing AR Process Simulation of Non-Gaussian Radar Clutter via Series-Based Analytic Continuation} 
\author{Xingxing~Liao, Junhao~Xie,~\IEEEmembership{Senior Member,~IEEE}

\thanks{This work was supported by the National Natural Science Foundation of China under Grant 62371162.} 
\thanks{The authors are with the Key Laboratory of Marine Environmental Monitoring and Information Processing, Harbin Institute of Technology, Harbin 150001, China. E-mail: (beehater552x@gmail.com; xj@hit.edu.cn). \textit{(Corresponding author: Junhao Xie.)}}}   

\maketitle
\begin{abstract} 
    Due to the conceptual simplicity, the linear filtering framework, notably the autoregressive (AR) process, has a long history in simulating clutter sequences with specified probability density functions (PDFs) and autocorrelation functions (ACFs). 
	However, linear filtering inevitably distorts the input distribution, which may lead to inaccurate PDF reproduction or restrict applicability to very simple ACFs. 
    To address these challenges, this study proposes a series-based analytic continuation strategy that revitalizes AR process clutter simulation by accurately precomputing the input pre-distortion required to compensate for AR filtering. 
    First, the moments and cumulants of the AR input are derived based on the input-output relationship of the AR process, facilitating the moment and cumulant expansions of the Laplace transform (LT) and the logarithmic LT around zero, respectively.  
    Second, both series expansions are analytically continued via the Pad\'{e} approximation (PA) to recover the LT over the full complex plane. 
    Notably, the PA-based continuation of the moment expansion, a conventional choice, can be highly inaccurate when the LT exhibits strong oscillations. 
    By contrast, given the logarithmic LT generally has a simpler structure, the continuation of the cumulant expansion provides a more stable and accurate alternative. 
    Third, the LT recovered from the cumulant expansion facilitates fast simulation of the AR input non-Gaussian white sequence via a random variable transformation method, thereby enabling an efficient AR process. 
	Finally, simulations demonstrate that the proposed strategy enables accurate and fast simulation of non-Gaussian correlated clutter sequences.  
\end{abstract}  

\begin{IEEEkeywords} 
	radar clutter simulation, linear filtering, autoregressive (AR) process, analytic continuation, Pad\'{e} approximation (PA).  
\end{IEEEkeywords} 

\section{Introduction} 
\label{sec:intro} 
\IEEEPARstart{A}{ccurate} radar clutter modeling plays an important role in analyzing the detection performance of maritime surveillance radars \cite{2013-Ward-seaclutter,2022-Rosenberg-seaclutter}.  
The compound Gaussian (CG) family of distributions has been widely used and verified to model real-world sea clutter under certain conditions, including the K distribution, Pareto distribution, and CG distribution with inverse Gaussian texture \cite{Xia2023-models}. 
Specifically, the recently proposed CG model with positive tempered $\alpha$-stable (PT$\alpha$S) texture can accurately fit measured sea clutter data with a wide range of tail behaviors, owing to its flexible tail-tuning capability \cite{Liao2024-PTaS}.

Good simulation of radar clutter is of great importance throughout the development and testing of radar systems \cite{2010Ward-simu}. 
Engineers rely on high-fidelity clutter simulations to evaluate algorithmic performance under controlled and repeatable conditions. 
For the CG clutter simulation, the most common practice is to perform simulation separately for the speckle and texture components. 
While the coherent complex Gaussian speckle can be generated with well-established methods, the simulation of the non-Gaussian correlated  positive texture component is considerably more challenging. 
Two typical frameworks for simulating a real-valued non-Gaussian process with prescribed probability density function (PDF) and autocorrelation function (ACF) or power spectral density (PSD) are the zero-memory nonlinearity (ZMNL) transform and the linear filtering, which are introduced below:  
\begin{itemize}
	\item \textit{ZMNL.} 
	It adopts a two-step simulation procedure: first, a correlated Gaussian process is generated; it is then passed through a nonlinear transformation to obtain the desired non-Gaussian correlated process \cite{1968Gujar-ZMNL}.  
	Since the nonlinear transformation for distribution control also distorts the correlation, a key step of this approach is to precompute the correlation of the underlying Gaussian process. 
	At present, ZMNL is the most widely used technique for simulating clutter texture, including Gamma texture (leading to K) and inverse Gamma (leading to Pareto) texture  
	\cite{Rosenberg2019-Models,Bocquet-2014-CG-InvGamma,Bocquet2013-Pareto, Wang2015-AP}.
	
	Nevertheless, for more complicated texture models (e.g., PT$\alpha$S), ZMNL becomes considerably more difficult to implement, as it relies on efficient inverse cumulative distribution function (CDF) algorithms or fast numerical evaluation of the associated transforms, which may not be unavailable. 
	Thus, alternative frameworks are worth exploring for simulating such clutter textures.

	\item \textit{Linear filtering.} 
	In contrast to ZMNL, an alternative two-step framework naturally exists: first, a non-Gaussian white sequence is generated; it is then passed through the linear filter to obtain the desired non-Gaussian correlated process. 
	Since linear filtering for correlation control also distorts the distribution, a key step of this approach is to precompute the distribution of the underlying non-Gaussian white sequence. 
    
	In early work, Oliver and Tough \cite{Oliver1986-ARsimu} proposed both exact and approximate methods for generating a correlated Gamma sequence by passing an uncorrelated Gamma-distributed sequence through a moving-average filter. 
	Specifically, the exact method produces Gamma sequences with a exponential ACF, whereas for more general ACFs, the approximate method preserves only the first- and second-order moments. 
	Szajnowski and Wynne \cite{Szajnowski2002-ARsimu} employed a first-order autoregressive (AR) process to simulate positive $\alpha$-stable (P$\alpha$S) sequences with a exponential ACF. This method is essentially equivalent to the exact approach in \cite{Oliver1986-ARsimu} and, in principle, is applicable to distributions possessing the infinite divisibility property. 
\end{itemize} 

Despite these early efforts, relatively little other attention has been paid to the linear filtering framework in non-Gaussian clutter simulation literature during the 21st century. 
Accurate simulation of non-Gaussian sequences with various ACFs under this framework remained a largely unresolved problem until the emergence of related studies in wind-resistant structural engineering \cite{Li2012-AR,Ma2019-new,Wu2024-AR-ME,Zhao2025-AR-ME,Wu2025-AR-MEM-PJTM}.

In \cite{Li2012-AR}, Li and Li first generated a white sequence with a prescribed distribution and then passed it through an AR process to impose the desired correlation. However, since AR filtering also alters the distribution, the resulting sequence typically deviates from the desired distribution.  
To address this issue, Ma \textit{et al.} \cite{Ma2019-new} derived the first four moments of the AR input from the input-output relationship. 
Using these moments, the conventional Johnson transformation was employed to generate the underlying non-Gaussian input sequence, expecting that the AR output matches both the prescribed PDF and ACF.  
Nevertheless, reliance on only four moments is generally insufficient to fully characterize the underlying PDF. 
To alleviate this limitation, the maximum entropy principle has been introduced to recover the input's PDF using a larger set of moments \cite{Wu2024-AR-ME,Zhao2025-AR-ME}, but this comes at the cost of computational efficiency. Moreover, the use of a large number of moments may lead to spurious oscillations or fluctuations in the PDF tail \cite{Zhao2025-AR-ME,Wu2025-AR-MEM-PJTM}.  
Therefore, it is of considerable interest to develop alternative techniques within the AR framework that enable accurate and efficient simulation of non-Gaussian correlated sequences.  

Motivated by these considerations, this study proposes a series-based analytic continuation strategy to revitalize AR process simulation of non-Gaussian processes.  
First, based on the input-output relationship of the AR process, the AR input's moments and cumulants are derived. These statistics are then used to construct the moment and cumulant expansions that represent the Laplace transform (LT) and logarithmic LT around zero, respectively. 
Second, noting that both series expansions contain the complete information of the LT \cite[Ch. V.8]{Gamelin2003-Ana-con}, we perform analytic continuation on each series to recover the LT over the entire complex plane. 
Specifically, following the conventional idea, the Pad\'{e} approximation (PA) \cite{Amindavar2002-pdfappro-Pade} is first applied to the moment expansion. 
However, for LTs exhibiting strong oscillatory attenuation, this may result in limited recovery. By contrast, the logarithmic LT typically has a simpler structure, making PA-based continuation of the cumulant expansion more stable and accurate. 
Once the input LT is obtained, the corresponding PDF can be recovered via the inverse LT (ILT). 
Numerical analysis demonstrates that the cumulant-expansion approach provides more stable and accurate LT and PDF recovery than the moment-expansion approach. 
Third, based on the recovered LT from the cumulant expansion, fast simulation of the AR input non-Gaussian white sequence is achieved. 
The complete AR process simulation procedure is detailed accordingly.   
Simulations for different PT$\alpha$S clutter textures validate that the cumulant expansion-based analytic continuation strategy is both accurate and efficient. 

The contributions of this study are summarized as follows: 
\begin{itemize}
	\item 
	We revitalize AR-based simulation of non-Gaussian processes via a series-based analytic continuation strategy, in which the PA-based continuation of the cumulant expansion in the logarithmic-LT domain enables accurate recovery of the AR input's LT and PDF. 
	\item 
	A fast AR input simulation scheme is developed by exploiting the multiplicative structure of the recovered LT through a random variable (RV) transformation method, avoiding the need for numerical CDF inversion or time-consuming rejection resampling. 
	\item 
	Compared with other methods within the linear filtering framework, such as Johnson transformation \cite{Ma2019-new} and maximum entropy-based approaches \cite{Wu2024-AR-ME,Zhao2025-AR-ME,Wu2025-AR-MEM-PJTM}, the proposed method more readily exploits a richer set of cumulants/moments to achieve accurate simulation with only simple operations. 
	Relative to the ZMNL framework, it is particularly well suited to models for which a closed-form inverse CDF is unavailable. 
\end{itemize} 

The remainder of this paper is organized as follows.
Section \ref{sec:review-clutter} reviews the CG clutter characteristics that need to be reproduced in simulation. 
Section \ref{sec:review-simumtds} briefly introduces the ZMNL method and the linear filtering via AR process for clutter simulation, and explains the motivation for adopting and revitalizing the AR-based simulation in this study.  
Section \ref{sec:Proposed-method} presents the proposed series-based analytic continuation strategy for accurate precomputation of the LT and PDF of the AR input. 
Based on Section \ref{sec:Proposed-method}, Section \ref{sec:Proposed-method-simu} designs the fast simulation of the AR input and details the complete procedure of the AR process simulation. 
Section \ref{sec:Proposed-method-verify} evaluates the simulation performance of the proposed analytic continuation strategy. 
Finally, Section \ref{sec:conclusion} concludes the paper.

\section{Characteristics of Sea Clutter} 
\label{sec:review-clutter}   
It is widely recognized that the CG family of distributions is the one of the most suitable statistical model for sea clutter \cite{Rosenberg2019-Models} and has achieved widespread adoption in radar detection design and analysis \cite{2016-Maio-seaclutter-Det}. 
Assuming negligible background noise,  the CG clutter model is represented as 
\begin{equation} 
	\begin{split} 
		X_\text{I}+jX_\text{Q}  = \sqrt{V} \left[Z_\text{I}+jZ_\text{Q}\right]  
		\label{eq-complex-clutter-CG} 
	\end{split} 
\end{equation} 
where $Z_\text{I}+jZ_\text{Q}$ and $V$ denote the complex Gaussian speckle and the non-negative texture. 
$Z_\text{I}+jZ_\text{Q}$ and $V$ are independent. 
The texture $V$ can be any suitable non-negative RV.   

\subsection{Amplitude Distribution}  
As indicated by (\ref{eq-complex-clutter-CG}), the statistical properties of the CG model are governed by the texture. 
Among existing texture models, the Gamma distribution is currently the most widely accepted and adopted in the literature. Other commonly used texture models include the inverse Gaussian, inverse Gamma, and P$\alpha$S distributions. 
Besides, the positive tempered $\alpha$-stable (PT$\alpha$S) distribution, a recently introduced  clutter texture model, including the gamma, inverse Gaussian, and P$\alpha$S as special cases. 

In the following, we briefly introduce the most widely used Gamma texture model and the state-of-the-art PT$\alpha$S texture model. 
\begin{itemize}
	\item \textit{gamma distribution}:  
	\begin{equation}  
		\begin{split}  
			f_V(v;\alpha_\text{gm},\lambda_\text{gm}) = \frac{1}{\Gamma(\alpha_\text{gm})\lambda_\text{gm}^{\alpha_\text{gm}}} v^{\alpha_\text{gm}-1} \exp \left( -\frac{v}{\lambda_\text{gm}}\right)
			\label{eq-gamma-pdf} 
		\end{split} 
	\end{equation}  
	where $\alpha_\text{gm}$ and $\lambda_\text{gm}$ are the shape and scale parameters, respectively.

	\item \textit{PT$\alpha$S distribution}:    
	For a PT$\alpha$S-distributed RV, its LT can be defined as   
	\begin{equation} 
		\begin{split} 
			\mathcal{L}_V(s) = \exp \left[ {\frac{\gamma }{{\alpha{\eta ^{\alpha}}}}\left( {1 - {{\left( {\eta s + 1} \right)}^{\alpha}}} \right)} \right]   
			\label{eq-model-Pos-LPT} 
		\end{split} 
	\end{equation}  
	where $\alpha\in(0,1]$, $\gamma>0$, and $\eta\in(0,+\infty)$ are the characteristic exponent, scale, and truncation parameter, respectively.   
	When $\eta \to \infty$, it reduce to the P$\alpha$S distribution's LT as $\exp \left[ - \gamma s^{\alpha}/\alpha \right]$. Thus, the PDF of PT$\alpha$S can be expressed as 
	\begin{equation} 
		\begin{split} 
			&f_V\left( v;\alpha,\gamma,\eta  \right) = \exp \left( { - \frac{v}{\eta}  + \frac{ \gamma}{\alpha \eta ^{\alpha}}} \right)f_\text{P$\alpha$S}\left( v;\alpha,\gamma \right) 
			\label{eq-model-Pos-LPT-pdf} 
		\end{split} 
	\end{equation} 
	where $f_\text{P$\alpha$S}\left( v;\alpha,\gamma \right)$ is the PDF of the P$\alpha$S distribution \cite{Aalo2013-PTaS} that can be computed by 
	\begin{equation} 
		\begin{split} 
		f_\text{P$\alpha$S}\left( v;\alpha,\gamma \right) &= {{\cal L}^{ - 1}}\left[ {\exp \left( { - {\gamma }{s^\alpha }/{\alpha }} \right)} \right]\left( v \right) \\
		&= {v^{ - 1}}H_{1,1}^{0,1}\left[ {\frac{\alpha }{\gamma }{v^\alpha }\left| {\begin{array}{*{20}{c}}
					{(1,1)}\\
					{(1,\alpha )}
			\end{array}} \right.} \right]
			\notag 
		\end{split} 
	\end{equation} 
	where ${\cal L}^{ - 1}$ donated the ILT, and $H:[\cdot | \cdot]$ donates the Fox's H-function \cite{Mathai2009-HFunction}. 
	Besides the P$\alpha$S distribution, when $\alpha \to 0$, (\ref{eq-model-Pos-LPT-pdf}) reduces to the gamma distribution in (\ref{eq-gamma-pdf}) with $\alpha_\text{gm}=\gamma$ and $\lambda_\text{gm}=\eta$; and when $\alpha = 0.5$, it reduces to the inverse Gaussian distribution. 
\end{itemize}

Let $Z_\text{I}$ and $Z_\text{Q}$ are mutually independent Gaussian RVs, $Z_\text{I}, Z_\text{Q} \sim \mathcal{N}(0,2)$, the PDF of the amplitude RV, $R=\sqrt{X_\text{I}^2+X_\text{Q}^2}$, is derived as \cite{Liao2024-PTaS} 
\begin{equation} 
	\begin{split} 
		f_R\left( r \right) =\int_0^\infty  {\frac{r} {2v}} \exp \left( { - \frac{r^2}{{4v}}} \right){f_V}(v)dv   
		\label{eq-Biv-NTS-PDF-R-CG} 
	\end{split} 
\end{equation} 
where ${f_V}(v)$ is the texture PDF, working as the mixing distribution. 

\begin{figure*}[t] 
	\centering
	\subfloat[]{ 
		\includegraphics[width=0.60\linewidth]{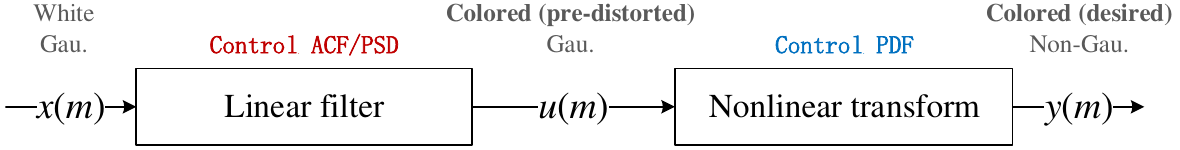} 
	} 
	
	\subfloat[]{ 
		\includegraphics[width=0.60\linewidth]{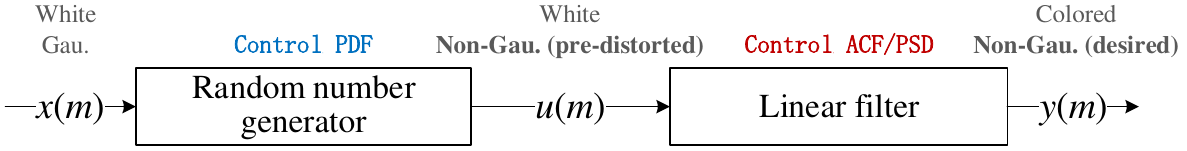}
	}	
	\caption{Comparison of the simulation procedure of two conventional real-valued sequence simulation frameworks. (a) ZMNL via correlation distortion; (b) linear filtering via distribution distortion.}  
	\label{fig-comparison-ZMNL-AR} 
\end{figure*}

\subsection{Temporal Correlation of the Speckle and Texture Components}   
The speckle component, related to the local surface geometry and scattering mechanisms, varies rapidly with a correlation time on the order of milliseconds. Short-term speckle correlation is typically modeled using a Gaussian-shaped Doppler spectrum \cite{Watts2012-EDSM}.  
On a longer time scale, under relatively calm sea conditions, the influence of long waves is typically moderate. However, at high sea states, the modulation of sea-surface Bragg scatterers by long waves leads to a periodic decay in the temporal correlation of the texture component. This behavior can be modeled using an exponential attenuation function modulated by a cosine term \cite{Bocquet2013-Pareto,Wang2015-AP}.

It is widely acknowledged that for CG clutter, while the coherent complex Gaussian speckle can be generated easily, simulating the texture—i.e., a non-Gaussian correlated process—is more challenging. ZMNL and linear filtering are two conventional frameworks for non-Gaussian correlated process simulation, as reviewed and compared below.

\section{Review of the ZMNL and the Linear Filtering via AR Process} 
\label{sec:review-simumtds}
Fig. \ref{fig-comparison-ZMNL-AR} compares the simulation procedure of the ZMNL method and the linear filtering. As shown, the two methods are complementary: ZMNL primarily relies on correlation distortion, whereas the linear filtering relies on distribution distortion. This fundamental difference leads to substantially different implementation procedure for the two approaches, as detailed below. 

\subsection{ZMNL}  
In Fig. \ref{fig-comparison-ZMNL-AR}(a), the ACF of the nonlinear transformation input should be recovered by backtracking from the prescribed output. 
To do this, ZMNL first designs the nonlinear transformation to match the desired distribution of $y(m)$. This transformation is then used to derive the correlation of $u(m)$. (Based on the results from the two steps, the design of the linear filter can be easily determined.)

\subsubsection{Inverse transform for desired output distribution}  
The zero-mean and unit-variance Gaussian distribution is mapped to the uniform distribution using the error function. The uniform distribution is then mapped to the desired PDF using its inverse CDF. 
For a sequence $\varsigma$ with a CDF $F_\text{dis}(\cdot)$, the MNLT is 
\begin{equation} 
	\begin{split} 
		\varsigma(x)=F_\text{dis}^{-1}(\operatorname{erfc}(x / \sqrt{2}) / 2)  
		\label{eq-ZNML-IT} 
	\end{split} 
\end{equation} 
where $\operatorname{erfc}(\cdot)$ is the complementary error function. 
In some cases, such as the exponential distribution, there is an analytic function for the inverse CDF.

\subsubsection{Correlation predistortion for desired output correlation}  
The correlation function of the output process $\varsigma$ is 
\begin{equation} 
	\begin{split} 
		&\langle \varsigma(0) \varsigma(t)\rangle = \\
		&\int_{-\infty }^{+\infty} \int_{-\infty }^{+\infty} \varsigma \left ( x_1 \right ) \varsigma \left ( x_2 \right ) P_U\left ( x_1,x_2,R_U\left ( t \right )   \right )  d x_1 d x_2
		\label{eq-ZNML-corr-2D-intergral} 
	\end{split} 
\end{equation} 
where 
\begin{equation} 
	\begin{split} 
		&P_U\left ( x(0),x(t),R_U\left ( t \right ) \right ) = \frac{1}{2\pi\sqrt{1-R_U^2\left ( t \right )}}  \\
		&\quad  \times \exp\left ( -\frac{x^2(0) +x^2(t)-2x(0)x(t)R_U(t)}{2\left ( 1-R_U^2(t) \right ) }  \right )   
		\label{eq-ZNML-corr-2D-intergral-RG} 
 	\end{split} 
\end{equation} 

The two-dimensional integration implicit in (\ref{eq-ZNML-corr-2D-intergral}) can be avoided by using
\begin{equation} 
	\begin{split} 
		&P_U\left ( x_1,x_2,R_U\left ( t \right ) \right ) = \frac{\exp \left ( -(x_1^2 +x_2^2)/2 \right ) }{2\pi} \\
		&\quad \times \sum_{n=0}^{\infty } \frac{H_n(x_1/\sqrt{2})H_n(x_2/\sqrt{2})}{2^n n!} R_U^n(n)  
	\end{split} 
\end{equation} 
where $H_n(\cdot)$ is the physicist's Hermite polynomial \cite{Weisstein2002-Hermite}. 

Then, the relationship between the desired correlation function and the correlation function $R_{U}(t)$ for the Gaussian process can be given by a series expansion in the Hermite polynomials $H_{n}(x)$: 
\begin{equation} 
	\begin{split} 
		&\langle \varsigma(0)\varsigma(t)\rangle=\frac{1}{\pi} \sum_{n=0}^{\infty} \frac{R_{U}^{n}(t)}{2^{n} n!} \times \\ 
		&\quad \left [ \int_{-\infty}^{\infty} \exp \left(-x^{2}\right) H_{n}(x) F_\text{dis}^{-1} \left(\frac{\operatorname{erfc}(x)}{2} \right)  d x  \right ]^2.   
		\label{eq-ZNML-corr} 
	\end{split} 
\end{equation} 

As indicated by the complicated appearance of this series, efficient algorithms for evaluating the inverse CDF, i.e., $F_{\mathrm{dis}}^{-1}(\cdot)$, are required for the practical implementation of ZMNL. Thanks to the development of fast algorithms for computing the inverse CDFs of the gamma and inverse gamma distributions, this step can be readily carried out numerically for these models \cite{Rosenberg2019-Models,Bocquet-2014-CG-InvGamma,Bocquet2013-Pareto}. 
In addition, the integrals in (\ref{eq-ZNML-corr}) can be efficiently evaluated using Gauss-Hermite quadrature with a modest number of points. 
While the series converges rapidly when the shape parameters are large, its convergence becomes slower as they decrease. 

More importantly, for more complicated texture models (such as the PT$\alpha$S distribution that provides more accurate fit but its PDF is available only in integral form \cite{Liao2024-PTaS}), ZMNL becomes considerably more difficult to implement, as neither efficient inverse-CDF algorithms nor fast numerical evaluation of the associated transforms are readily available. 
Therefore, alternative simulation frameworks are worth exploring for modeling such clutter textures. This consideration motivates our investigation of the linear filtering framework in the following.

\subsection{Linear Filtering via AR process} 
Similar to ZMNL, we need to infer the input information of the linear filter from the prescribed output in a reverse manner. 
As shown in Fig. \ref{fig-comparison-ZMNL-AR}(b), the linear filtering is first designed to match the desired correlation of $y(m)$. This linear transformation is then used to derive the distribution of $u(m)$. (Based on the results from the two steps, the design of well-established random number generators, such as nonlinear transformation and rejection sampling, can be  specified.)

\subsubsection{AR process for desired output correlation}  
AR and autoregressive moving average (ARMA) processes are commonly used to approximate any arbitrary ACF structure of a stochastic process. They are typically expressed as 
\begin{equation} 
	\begin{split}  
		y(m) = -\sum_{k=1}^{p} a_{k} y(m-k)+u(m)  
		\label{eq-AR} 
	\end{split} 
\end{equation} 
\begin{equation} 
	\begin{split} 
		y(m) = -\sum_{k=1}^{p} a_{k} y(m-k)+\sum_{l=0}^{q} b_{l} u(m-l)  
		\label{eq-ARMA} 
	\end{split} 
\end{equation} 
where $y(m)$ and $u(m)$ are the output process and the white input noise with zero mean, respectively.
$p$ and $q$ are the orders of the AR component and the moving average component, respectively. 
$a_k$ and $b_l$ are the process coefficients. 
Given the prescribed ACF, Yule-Walker method can be used to compute $a_k$ and $b_l$ \cite[Chapter 3]{Box2008-book}. 
In particular, an AR process with a sufficiently high order can achieve simulation accuracy comparable to that of an ARMA process, of arbitrary order. Hence, to simplify the discussion and without loss of generality, we focus on the AR process in the following analysis.

The AR process given in the form of (\ref{eq-AR}) is widely adopted in the literature. 
However, this process may not be convenient for analysis techniques that rely on independence. 
Take the AR(2) process (i.e., $y(m) = - a_1 y(m-1)- a_2 y(m-2) + u(m)$ for example, the input elements $y(m-1)$ and $y(m-2)$ are correlated. 
Fortunately, the output of AR process can be transformed into 
\begin{equation} 
	\begin{split} 
		y(m) = \sum_{i=0}^{\infty} h_{i} u(m-i) \simeq \sum_{i=0}^{L_\text{IR}} h_{i} u(m-i)   
		\label{eq-AR-ARMA-Green} 
	\end{split} 
\end{equation} 
where $h_{i}$ is the unit impulse response. 
Theoretically, $h_{i}$ is an infinitely long sequence, whose magnitude approaches zero as $i$ increases for stable systems. The decay rate of the impulse response is determined by the transfer function of the underlying linear system. 
For weakly correlated sequences, the impulse response decays rapidly, whereas for strongly correlated sequences, it exhibits a slower decay. 
$L_\text{IR}$ denotes the finite lag at which the infinite summation is truncated, chosen such that $|h_i|$ falls below a predefined threshold (e.g., $|h_i| < 10^{-3}$), ensuring that the truncation error is negligible in practice.

\subsubsection{Distribution predistortion for desired output distribution}  
To ensure that the AR output achieves both the desired ACF and PDF, the distribution of the driving sequence $u(m)$ must be appropriately pre-distorted. 
To this end, typical studies derive the first several moment  relationships between the input and output processes via \cite{Ma2019-new,Wu2024-AR-ME,Zhao2025-AR-ME} 
\begin{equation} 
	\begin{split} 
         \mathbb{E} \left\{ {{{\left[ {y(m)} \right]}^n}} \right\} = \mathbb{E} \left\{ {{{\left[ {\sum_{i = 0}^{L_\text{IR}}  {{h_i}} u(m - i)} \right]}^n}} \right\}, n = 1,2, \cdots 
		\label{eq-AR-ARMA-moment-rela} 
	\end{split} 
\end{equation} 
where $\mathbb{E} \left\{ \cdot \right\} $ donates the expectation operator.   

Then, by matching the first four moments using the conventional Johnson transformation \cite{Ma2019-new} or a larger set of moments using the maximum entropy principle \cite{Wu2024-AR-ME,Zhao2025-AR-ME}, the non-Gaussian sequence $u(m)$ can be generated. 
However, the former is generally insufficient to fully characterize the PDF, while the latter suffers from a trade-off between computational efficiency and accuracy and may converge to local optima, resulting in spurious tail oscillations.

To enable accurate and fast non-Gaussian correlated sequence simulation, we propose a series-based analytic continuation method to determine the distribution of the linear filter's input $u(m)$, as detailed below.

\section{Proposed Series-Based Analytic Continuation for AR input's Statistics} 
\label{sec:Proposed-method}   
This section proposes a series-based analytic continuation strategy for accurate and fast simulation of complicated non-Gaussian correlated processes. 
First, the moment expansion of the AR input's LT and the cumulant expansion of the logarithmic LT are constructed, respectively.  
Second, the PA is used to analytically continue the two series expansions to obtain the LT and PDF of the AR input. 
Finally, the performance of the PA-based continuation of the two different series expansions are compared.

\subsection{Series Expansion of AR Input's LT} 
Let $Y$ and $U$ denote the RVs corresponding to $y$ and $u$ in (\ref{eq-AR-ARMA-Green}), respectively. 
The LT relationship between $Y$ and $U$ can be expressed as
\begin{equation} 
	\begin{split} 
		\mathcal{L}_Y(s) = \prod_{i=0}^{\infty} \mathcal{L}_U(h_{i} s) \simeq \prod_{i=0}^{L_\text{IR}} \mathcal{L}_U(h_{i} s).    
		\label{eq-AR-LP} 
	\end{split} 
\end{equation}

Based on (\ref{eq-AR-LP}), the relationship between the $n$-th cumulants of $Y$ and $U$ can be expressed as  
\begin{equation} 
	\begin{split} 
		 \kappa_{Y,n} &= \left.(-1)^{n} \frac{d^{n}}{d s^{n}} \ln \mathcal{L}_Y(s)\right|_{s=0} \\
		 &=  \sum_{i=0}^{L_\text{IR}}  \left.(-1)^{n} \frac{d^{n}}{d s^{n}} \ln \mathcal{L}_U(h_{i}s)\right|_{s=0} \\
		 &=\sum_{i=0}^{L_\text{IR}}  h_{i}^n \kappa_{U,n}.   
		\label{eq-FT-Ray-var-y-cum} 
	\end{split} 
\end{equation} 

From (\ref{eq-FT-Ray-var-y-cum}), the cumulants of the AR input $U$ can be expressed in terms of those of the output $Y$, i.e., 
\begin{equation}  
	\begin{split} 
		\kappa_{U,n} =  \kappa_{Y,n}/\iota_n  
		\label{eq-AR-LP-cum-rela} 
	\end{split} 
\end{equation} 
where $\iota_n={\sum\nolimits_{i = 0}^{L_\text{IR}} {{h^n_i}}}$. 
Then, the moments of $U$ can be obtained from its cumulants as \cite{Withers2009moments} 
\begin{equation} 
	\begin{split} 
	     M_{U,n}= B_{n}\left(\kappa_{U,1}, \kappa_{U,2},\ldots, \kappa_{U,n}\right) 
		\label{eq-FT-Ray-var-u-mom} 
	\end{split} 
\end{equation} 
where $B_{n}(\cdot)$ is the $n$-th complete Bell polynomial \cite{Kim2021-Bell}. 

Given the cumulants/moments of a distribution, the corresponding LT can be approximated via a series expansion for small values. 
Specifically, the Taylor expansions of $U$'s LT and logarithmic LT around zero yield the following power series in terms of the moments $M_{U,n}$ and cumulants $\kappa_{U,n}$, respectively: 
\begin{equation}  
	\begin{split}   
		{\mathcal{L}_U}(s) &= \sum\limits_{n = 0}^\infty    \frac{{M_{U,n}}}{{n!}}  {{(-s)}^n}, \enspace s \to 0^+,     
		\label{eq-AR-U-cum-LP-series} 
	\end{split} 
\end{equation} 
\begin{equation}  
	\begin{split} 
		\log {\mathcal{L}_U}(s) &= \sum\limits_{n = 1}^\infty    \frac{{\kappa_{U,n}}}{{n!}}  {{(-s)}^n}, \enspace s \to 0^+.  
		\label{eq-AR-U-cum-logLP-series} 
	\end{split} 
\end{equation} 

These series typically encounter divergence issues. 
For example, the convergence radius of the series (\ref{eq-AR-U-cum-LP-series}) is \cite[3.39 Theorem]{Rudin1976-ratio}  
\begin{equation}  
	\begin{split}  
		R_\text{con} = 1/c_\text{lim}, \enspace \text{with } c_\text{lim} = \lim_{n\to \infty}  \left | {M_{U,n}}/{n!} \right |^{1/n}. 
		\label{eq-series-ratio-radi} 
	\end{split} 
\end{equation}  
If $c_\text{lim}=0$, $R_\text{con} = \infty$; whereas if $c_\text{lim}=\infty$, then $R_\text{con} = 0$. Then, the series in (\ref{eq-AR-U-cum-LP-series}) converges if $|s|<R_\text{con}$, and diverges if $|s|>R_\text{con}$. 
Take $U$ as a gamma variable for example to analyze the convergence of the series in (\ref{eq-AR-U-cum-LP-series}).  
The gamma PDF is given in (\ref{eq-gamma-pdf}), and the corresponding $n$-th moment is $\lambda_\text{gm}^n\Gamma(n+\alpha_\text{gm})/\Gamma(\alpha_\text{gm})$. Thus, the convergence radius is
\begin{equation}  
	\begin{split}  
    	R_\text{con} = 1/ \lim_{n\to \infty} \left ( \frac{\lambda_\text{gm}^n}{\Gamma(\alpha_\text{gm})} n^{\alpha_\text{gm}-1} \right )^{1/n} =  1/\lambda_\text{gm}. 
		\label{eq-series-ratio-gamma-radi} 
	\end{split} 
\end{equation}  

Thus, the series in (\ref{eq-AR-U-cum-LP-series}) converges for $|s|<1/\lambda_\text{gm}$; whereas it diverges for $|s|<1/\lambda_\text{gm}$. 
This implies that directly evaluating the series in (\ref{eq-AR-U-cum-LP-series}) to obtain ${\mathcal{L}_U}(s)$ is impractical for large $|s|$ values. 
More importantly, even within its region of convergence, the series—being a Taylor expansion—typically provides high accuracy only in a neighborhood of the expansion point. 
Consequently, alternative techniques are required to obtain reliable approximations beyond the local neighborhood of the expansion point.

Analytic continuation is a powerful tool for extending locally convergent series to a broader domain \cite[Ch. V.8]{Gamelin2003-Ana-con}. This follows from a fundamental result in complex analysis: all properties of a function analytic at a point are completely determined by its power series expansion at that point. Consequently, analytic continuation has played an important role in approximation theory and related fields. 

The PA is a representative analytic continuation technique \cite{Bender1999-Pade,Baker1996-Pade}. 
At present, when the Taylor expansion of a function—representing its local or asymptotic behavior near zero—is known, the PA is widely used to extend the series representation and recover the function’s global behavior  \cite{El2002-Pade,Tourandaz2017-Pade,Franceschini1985-borel-pade,2017Lvarez-borel-pade}. 
Next, we employ it to perform the analytic continuation to recover the statistics of the AR input $U$.

\subsection{Series-based Analytic Continuation for AR Input's LT \& PDF} 
This part separately uses the PA to analytically continue the moment expansion in (\ref{eq-AR-U-cum-LP-series}) and the cumulant expansion in (\ref{eq-AR-U-cum-logLP-series}), in order to obtain the LT and PDF of the AR input $U$.  

\subsubsection{Analytic continuation of moment expansion (\ref{eq-AR-U-cum-LP-series}): conventional way} 
Using the PA to a moment expansion to obtain the LT over the entire complex domain is a conventional approach \cite{Amindavar2002-pdfappro-Pade}. 
To obtain the LT of the AR input $U$, we first follow this idea and directly apply the PA to (\ref{eq-AR-U-cum-LP-series}): 
\begin{equation}  
	\begin{split} 
		  {[K,L]}_{\mathcal L}( s )  & = \frac{P(s)}{Q(s)} \simeq \sum\limits_{n = 0}^\infty    \frac{M_{U,n}}{n!}  {{(-s)}^n} 
		\label{eq-pade-1} 
	\end{split} 
\end{equation} 
where $P(s)$ and $Q(s)$ are polynomials of order $K$ and $L$ respectively, i.e.,
\begin{equation}  
	\begin{split} 
		& P(s) =  \sum\limits_{i = 0}^K  P_i s^i, \enspace Q(s) =  \sum\limits_{j = 0}^L  Q_j s^j
		\label{eq-pade-2} 
	\end{split} 
\end{equation} 
where $P_i$ and $Q_j$ are constant coefficients determined by matching ${[K,L]}_{\mathcal L} (s)$ with $\mathcal{L}_U(s)$, via a straightforward matrix-based linear algebraic procedure, involving only the solution of several small-scale linear systems. 
For the sake of readability, the detailed solution procedure is reviewed in Appendix A.

Note that, in this study, we only consider the diagonal ($K=L$) and sub-diagonal ($K=L-1$) PAs because they play a particular role in the convergence of the Pad\'{e} sequence. 
Particularly, they provide upper and lower bounds to any Stieltjes series and if the problem is determinate they converge to the unique solution \cite{Bender1999-Pade}. 
Once $P_i$ and $Q_j$ are obtained, (\ref{eq-pade-1}) can be rewritten via partial fraction decomposition as
\begin{equation}  
	\begin{split} 
		{\mathcal{L}_U}( s ) \simeq {[K,L]}_{\mathcal L}( s ) & = \frac{ \sum\nolimits_{i = 0}^K  P_i s^i}{\sum\nolimits_{j = 0}^L  Q_j s^j}  = \sum\limits_{j = 1}^{L} \frac{\lambda_j}{s+a_j}, \enspace s \in \mathbb{C} 
		\label{eq-pade-3} 
	\end{split} 
\end{equation} 
where $a_i$ denotes the negative of the corresponding pole, and $\lambda_i$ denotes the associated residue. 
Note that if any $a_j$ is negative (corresponding to a pole in the positive half-plane), it clearly violates the correct analytic structure and should be discarded an unacceptable approximation \cite{Baker1996-Pade}. 

By applying the LT pair ${\mathcal{L}^{-1}} \left[  {1}/{(u+a)} \right] (s) = \exp (-as)$, we can take the ILT of (\ref{eq-pade-3}) and get the PDF of $U$ as  
\begin{equation}  
	\begin{split} 
		f_U(u) & = \sum\limits_{j = 1}^{L} \lambda_j \exp (-a_j u), \enspace  x>0   
		\label{eq-pade-4} 
	\end{split} 
\end{equation} 
where $u$ denotes the realization of $U$. 

\begin{figure}[t] 
	\centering
	\subfloat[]{ 
		\includegraphics[width=0.48\linewidth]{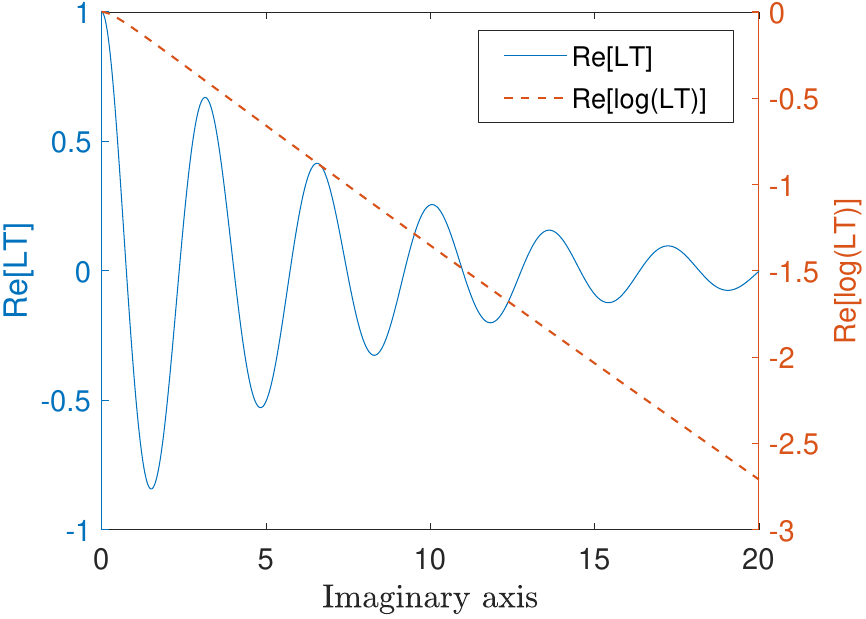} 
	} 
	\subfloat[]{ 
		\includegraphics[width=0.48\linewidth]{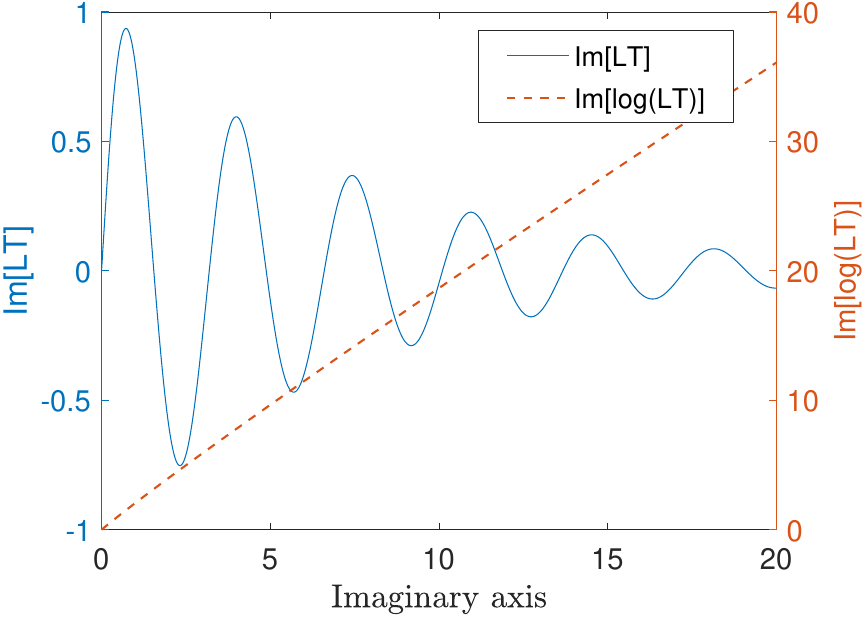} 
	}	
	\caption{Comparison of the LT and logarithmic LT of the PT$\alpha$S distribution with parameters $\alpha=0.95$, $\gamma = 2$, $\eta=4$. (a) Real part; (b) Imaginary part.}    
	\label{fig-dist-theo-LT} 
\end{figure}

\subsubsection{Analytic continuation of cumulant Expansion (\ref{eq-AR-U-cum-logLP-series}): new way}      
When the complex-valued LT exhibits strong oscillatory attenuation, the PA of the moment expansion (\ref{eq-AR-U-cum-LP-series}) easily fails to accurately recover the LT over the entire complex plane, as will be illustrated in Section \ref{sec:Proposed-method}-C.  
To address this limitation, we instead propose applying the PA to analytically continue the cumulant expansion (\ref{eq-AR-U-cum-logLP-series}), which corresponds to the logarithmic LT exhibiting a simpler functional structure (see the PT$\alpha$S example shown in Fig. \ref{fig-dist-theo-LT}). 

For (\ref{eq-AR-U-cum-logLP-series}), a re-indexing is first performed to facilitate the application of the PA: 
\begin{equation}  
	\begin{split} 
		\log {\mathcal{L}_U}(s) &=  -s\sum\limits_{n = 0}^\infty    \frac{{\kappa_{U,n+1}}}{{(n+1)!}}  {{(-s)}^n}, \enspace s \to 0^+.  
		\label{eq-AR-U-cum-logLP-series-re} 
	\end{split} 
\end{equation} 

Then, through the PA given in Appendix A, the series (\ref{eq-AR-U-cum-logLP-series-re}) of order $K + L$ are written as  
\begin{equation}  
	\begin{split} 
		{[K,L]}_{\log \mathcal L} \left( s \right) & = \sum\limits_{j = 1}^{L} \frac{-\lambda_j s}{s+a_j} 
		\label{eq-AC-Cum-logLT} 
	\end{split} 
\end{equation}  
where $a_i$ denotes the negative of the corresponding pole, and $\lambda_i$ denotes the associated residue. To ensure the correct analytic structure, any negative $a_j$ is excluded. 

Applying the exponential operation to (\ref{eq-AC-Cum-logLT}) gives the approximated LT as 
\begin{equation}  
	\begin{split} 
		&{\mathcal{L}_U}( s ) \simeq \exp \left ( {[K,L]}_{\log \mathcal L} \left( s \right)  \right )   \\
		&= \exp \left ( \sum\limits_{j = 1}^{L} \frac{-\lambda_j s}{s+a_j} \right ) = \prod_{j=1}^{L} \exp \left (  \frac{-\lambda_j s}{s+a_j} \right ) , \enspace s \in \mathbb{C^+}.   
		\label{eq-AC-Cum-LT} 
	\end{split} 
\end{equation} 

Therefore, $U$ can be expressed as the sum of independent RVs as 
\begin{equation}  
	\begin{split} 
		& U = \sum^{L}_{j=1} Z_j, \enspace \text{with }  {\mathcal{L}_{Z_j}}( s ) = \exp \left (  \frac{-\lambda_j s}{s+a_j} \right )
		\label{eq-simu-U-add}  
	\end{split} 
\end{equation} 
where $Z_j$, $j=1,2,\cdots, L$, are independent.

Accordingly, the PDF of $U$ satisfies 
\begin{equation}  
	\begin{split} 
		&f_U(u)= f_{Z_1}(u) \ast f_{Z_2}(u)  \ast \cdots  \ast f_{Z_L}(u)
		\label{eq-AC-Cum-LT-pdf} 
	\end{split} 
\end{equation} 
where $\ast$ donates the convolution symbol. Using the LT pair given in \cite[p.255, 5.66]{Oberhettinger-book} and performing some manipulation, we can get  
\begin{equation} 
	\begin{split} 
		&f_{Z_j}(u) = \mathcal{L}^{-1}\left[ \exp \left (  \frac{-\lambda_j s}{s+a_j} \right ) \right](u)  \\
		& = e^{-\lambda_{i}} \delta(u)+e^{-a_{i} u-\lambda_{i}} \sqrt{\frac{a_{i} \lambda_{i}}{u}} I_{1}\left(2 \sqrt{a_{i} \lambda_{i} u}\right) 
		\label{eq-AC-Cum-LT-pdf-Zj} 
	\end{split} 
\end{equation} 
where $j=1,2,\cdots, L$, $\delta(\cdot)$ denotes the Dirac delta function, and $I_{1}(\cdot)$ denotes the first-order modified Bessel function of the first kind.  

As indicated by (\ref{eq-AC-Cum-LT-pdf}) and (\ref{eq-AC-Cum-LT-pdf-Zj}), the ILT of (\ref{eq-AC-Cum-LT}) is difficult to be obtained in closed form. 
Given the complexity of the convolution operation, the PDF of $U$ can, if needed, be efficiently computed using a fast Fourier transform-based inversion method \cite{Josso2012-ILP}  
\begin{equation}  
	\begin{split} 
		f_U(u) &= \mathcal{L}^{-1}\left[ \prod_{j=1}^{L} \exp \left (  \frac{-\lambda_j s}{s+a_j} \right ) \right](u) 
		\label{eq-AC-Cum-pdf-num} 
	\end{split} 
\end{equation} 
where $\mathcal{L}^{-1}$ is performed using the fast Fourier transform.

\subsection{Comparison Between Analytic Continuation of Different Series}  
To compare the LT recovery performance using the PA-based continuations of different series (i.e., moment expansion in (\ref{eq-AR-U-cum-LP-series}) and cumulant expansion in (\ref{eq-AR-U-cum-logLP-series})), we consider $U$ following the gamma and PT$\alpha$S distributions, defined in (\ref{eq-gamma-pdf}) and (\ref{eq-model-Pos-LPT-pdf}), as two representative examples. Their $n$-th cumulant are $\kappa_{U,n} = \alpha_\text{gm} \lambda_\text{gm}^{n} \Gamma(n)$ and $\kappa_{U,n} = \gamma \eta^{n-\alpha} \Gamma(n-\alpha)/\Gamma(1-\alpha)$, respectively.    
The parameters of the gamma distribution are set as: $\alpha_\text{gm}=2$ and $\lambda_\text{gm}=1$; those of the PT$\alpha$S distribution are set as: $\alpha=0.95$, $\gamma = 2$, $\eta=4$, which empirically corresponds to a heavy-tailed clutter case.  
The PA order is set as $K=16$ and $L=17$. 

Note that, at this stage, we directly assume the PDF of the AR input $U$, rather than constructing $U$ via an AR process to match a prescribed output $Y$. This is because the theoretical PDF of $U$ is generally intractable in the latter case, which makes a quantitative comparison of the recovery performance inconvenient.

\subsubsection{Recovery performance of $U$'s statistics}  
\begin{figure}[t] 
	\centering
	\subfloat[]{ 
		\includegraphics[width=0.475\linewidth]{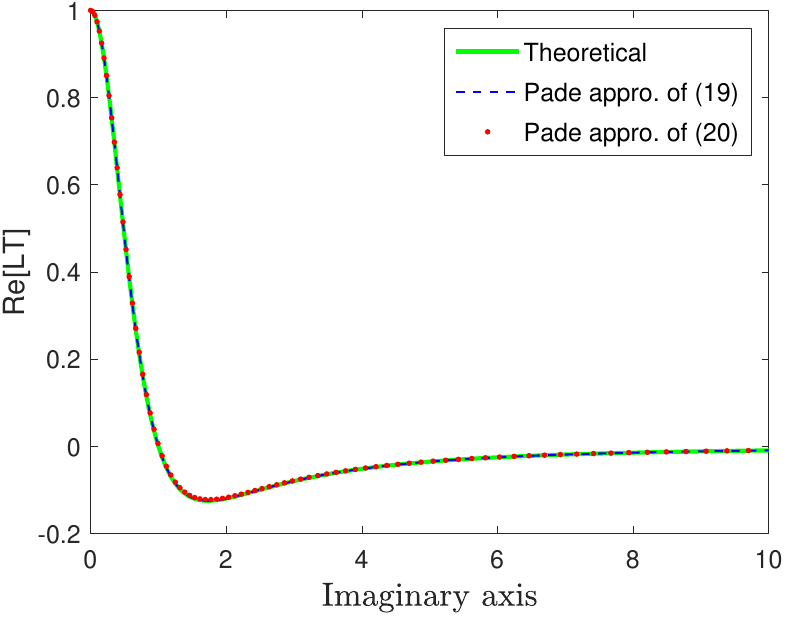} 
	} 
	\subfloat[]{ 
		\includegraphics[width=0.47\linewidth]{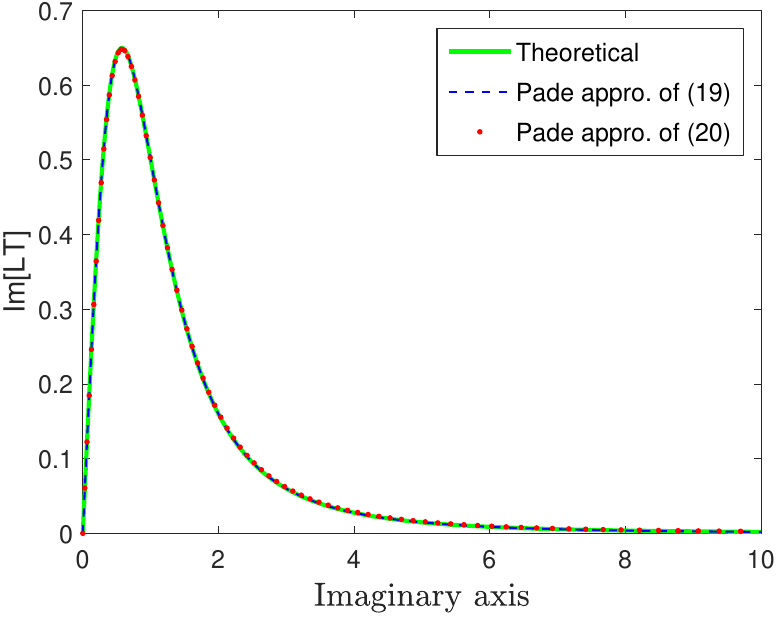}
	}	
	\caption{Comparison of the gamma LT recovery performance via the PA-based continuations of different series. (a) Real part of the LTs; (b) Imaginary part of the LTs.}  
	\label{fig-diff-Pade-LT-GM} 
\end{figure}

\begin{figure}[t] 
	\centering
	\subfloat[]{ 
		\includegraphics[width=0.475\linewidth]{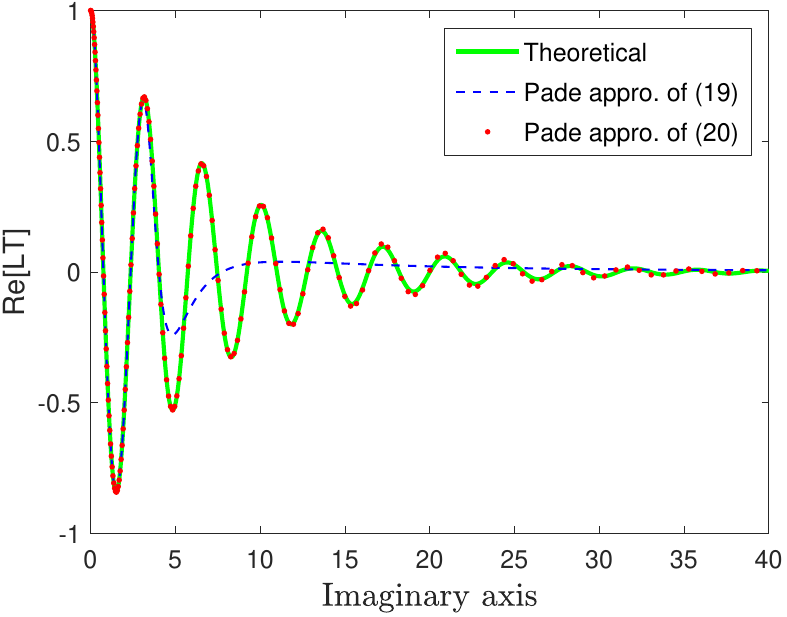} 
	} 
	\subfloat[]{ 
		\includegraphics[width=0.475\linewidth]{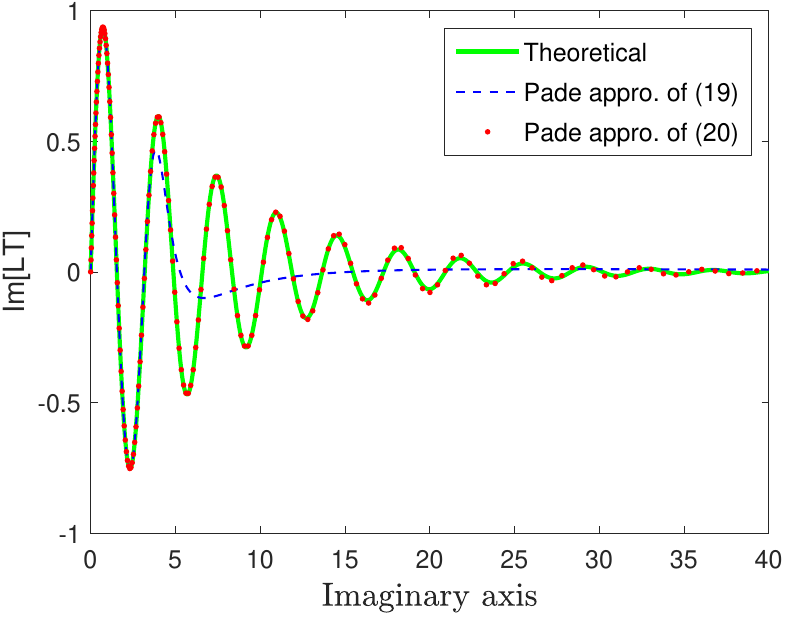}
	}	
	\caption{Comparison of the PT$\alpha$S LT recovery performance via the PA-based continuations of different series. (a) Real part of the LTs; (b) Imaginary part of the LTs.}  
	\label{fig-diff-Pade-LT} 
\end{figure}

Fig. \ref{fig-diff-Pade-LT-GM} compares the LT recovery performance for the gamma distribution using the PA-based continuations of different series, namely the moment expansion in (\ref{eq-AR-U-cum-LP-series}) and the cumulant expansion in (\ref{eq-AR-U-cum-logLP-series}).
Similarly, Fig. \ref{fig-diff-Pade-LT} presents the corresponding comparison for the PT$\alpha$S distribution.
It can be observed that, for the gamma distribution whose complex-valued LT exhibits relatively mild oscillations, the PA-based continuations of both series achieve satisfactory recovery performance. In contrast, for the PT$\alpha$S distribution with a complex-valued LT of strong oscillation attenuation, only the PA-based continuation of the cumulant expansion maintains good recovery accuracy. 
This behavior can be attributed to the fact that the moment and cumulant expansion series directly approximate the LT and the logarithmic LT, respectively. The logarithmic LT often has a simpler functional structure (see, e.g., the PT$\alpha$S example in Fig. \ref{fig-dist-theo-LT}), which is more amenable to accurate rational approximation.
Consequently, the PA-based continuation of the moment expansion performs well for the gamma distribution, but for the PT$\alpha$S distribution it only provides reliable results on imaginary axis less than 5, beyond which the recovery accuracy deteriorates rapidly. 

\begin{figure}[t] 
	\centering
	\subfloat[]{ 
		\includegraphics[width=0.471\linewidth]{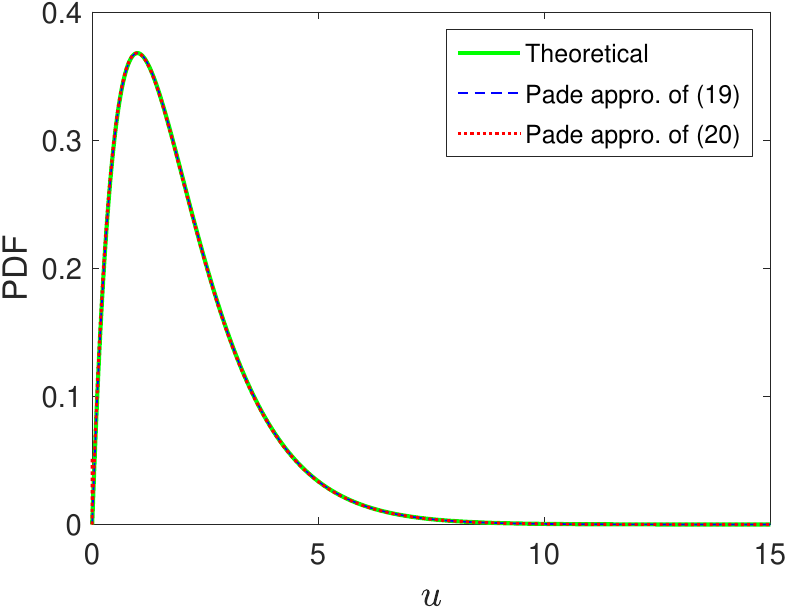} 
	} 
	\subfloat[]{ 
		\includegraphics[width=0.478\linewidth]{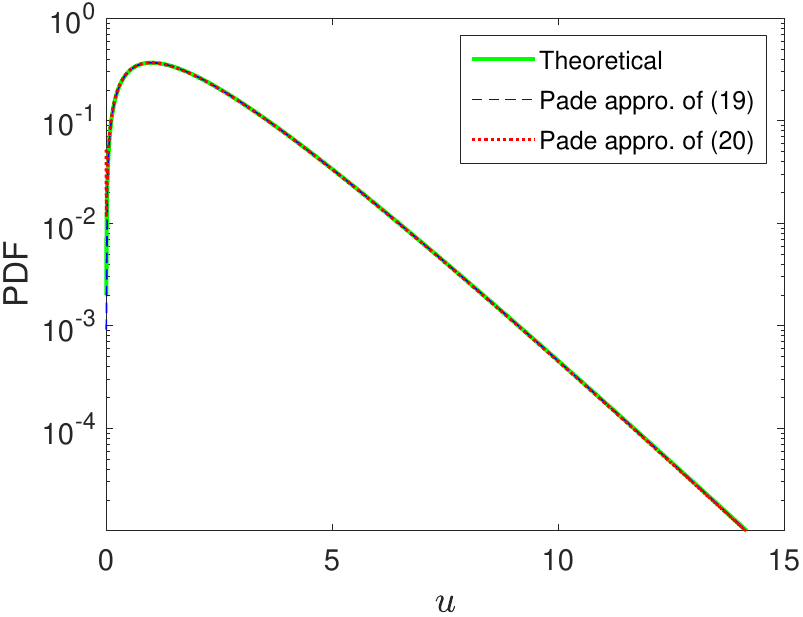}
	}	
	\caption{Comparison of the gamma PDF recovery performance via the PA-based continuations of different series. (a) Linear scale; (b) Logarithmic scale.}  
	\label{fig-diff-Pade-PDF-GM} 
\end{figure}

\begin{figure}[t] 
	\centering
	\subfloat[]{ 
		\includegraphics[width=0.46\linewidth]{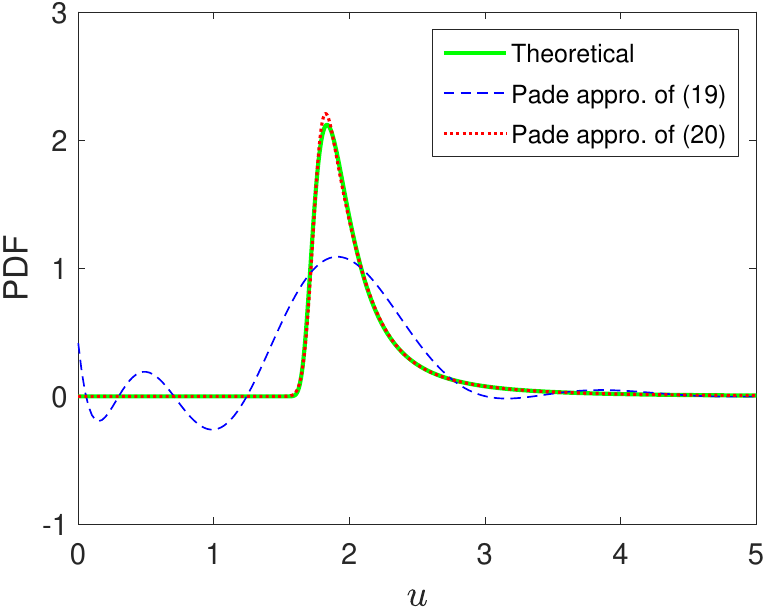} 
	} 
	\subfloat[]{ 
		\includegraphics[width=0.482\linewidth]{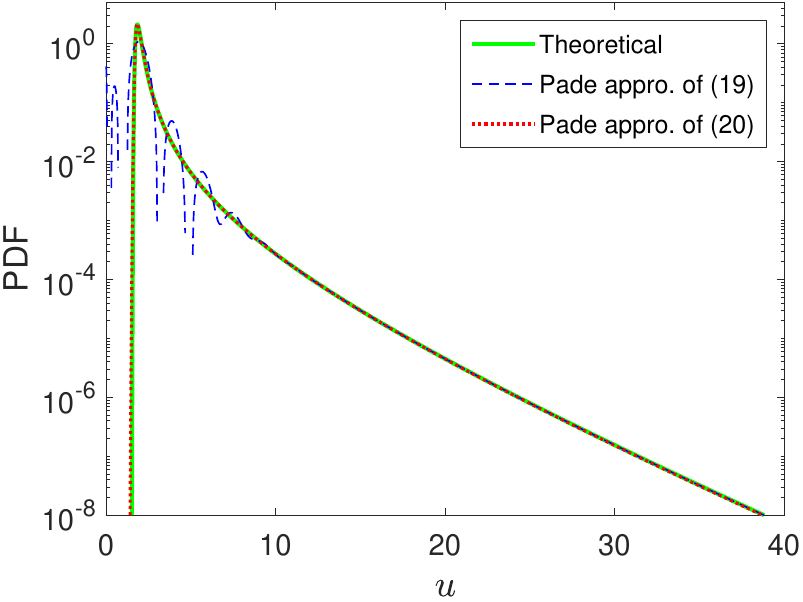}
	}	
	\caption{Comparison of the PT$\alpha$S PDF recovery performance via the PA-based continuations of different series. (a) Linear scale; (b) Logarithmic scale.}  
	\label{fig-diff-Pade-PDF} 
\end{figure}

The PDF recovering performance corresponding to Figs. \ref{fig-diff-Pade-LT-GM} and \ref{fig-diff-Pade-LT} is shown in Figs. \ref{fig-diff-Pade-PDF-GM} and \ref{fig-diff-Pade-PDF}, respectively. 
As can be seen, the PDF recovery performance is consistent with the corresponding LT reconstruction results. 
Specifically, for the gamma distribution, the PA-based continuations of both series achieve satisfactory PDF recovery performance . 
For the PT$\alpha$S distribution, the PA-based continuation of the moment expansion provides an accurate fit only in the extreme tail region (PDF below $10^{-3}$), while that based on the cumulant expansion yields accurate recovery in both the body and tail regions. 

The above analysis indicates that adopting the PA-based continuation to the cumulant expansion yields more accurate LT and PDF recovery performance than that based on the moment expansion series. Moreover, this strategy also offers advantages for fast AR process simulation, requiring only simple operations, as will be detailed in Section \ref{sec:Proposed-method-simu}.

\begin{table}[t] 
	\renewcommand\arraystretch{1.5} 
	\begin{center}
		\caption{Computation complexity of the PA-based continuations of different series for recovering the AR input $U$'s statistics}  
		\label{Table-Computation complexity}  
		\begin{threeparttable} 
			\begin{tabular}{p{3.2cm}<{\centering}|p{1.85cm}<{\centering}|p{2.0cm}<{\centering}} 
				\toprule 
				Series & Recovering $U$'s LT &  Recovering $U$'s PDF \\ 
				\midrule 
				Moment expansion in (\ref{eq-AR-U-cum-LP-series})           & $\mathcal{O}(L^3)$  &  $\mathcal{O}(L_x)$ \\  
				\cline{1-3} 
				Cumulant expansion in (\ref{eq-AR-U-cum-logLP-series})  & $\mathcal{O}(L^3)$ &  $\mathcal{O}(L_x L_s \log L_s)$ \\  
				\bottomrule  
			\end{tabular} 
		\end{threeparttable} 
	\end{center} 
\end{table}

\subsubsection{Computational complexity} 
For the PA-based continuations of the moment and cumulant expansions to get the LT of $U$, the dominant computational cost arises from solving the linear system in (\ref{eq-A4-1}), which involves the inverse of a dense matrix of size $L\times L$, leading to a computational complexity of $\mathcal{O}(L^3)$. 

As the PA-based continuation of the moment expansion yield closed-form PDF in (\ref{eq-pade-4}), the PDF evaluation involves only pointwise computations, resulting in a computational complexity of $\mathcal{O}(L_x)$, where $L_x$ is the number of PDF samples. 
In contrast, it is hard for the PA-based continuation of the cumulant expansion to get the closed-form PDF expression from the ILT in (\ref{eq-AC-Cum-LT-pdf-Zj}), a numerical inversion via the fast Fourier transform is required. Consequently, the computational complexity is $\mathcal{O}(L_x L_s \log L_s)$, where $L_s$ denotes the number of LT samples.

To summarize, the PA-based continuations of both series exhibit moderate computational complexity, as the order $L$ is small (typically less than 20). 
When PDF evaluation is required, the PA-based continuation of the cumulant expansion incurs a higher computational cost due to the need for numerical PDF computation; however, this cost is mild with the aid of the fast Fourier transform.

For both PDFs recovered from the moment and cumulant expansions, the inverses of their CDFs cannot be expressed in closed form. 
In such cases, general random number generation techniques, such as the composition method or rejection sampling, can be employed to simulate the sequence; however, these methods are computationally inefficient and impractical for real-time implementation.  

It is noteworthy that these issues can be avoided in our AR process simulation based on the cumulant expansion. 
Specifically, we do not need to compute the AR input's PDF; instead, by exploiting the multiplicative structure of the LT in (\ref{eq-AC-Cum-LT}), fast and accurate simulation can be achieved, as detailed below.   

\section{Fast AR Process Simulation}  
\label{sec:Proposed-method-simu}  
In the above, we have shown that the PA-based continuation of the cumulant expansion provides accurate recovery of the AR input's LT, which underpins the fast AR process simulation developed based on a RV transformation method in this section. 

\subsection{Simulation of AR Input Non-Gaussian White Sequence} 
The PA-based continuation of the cumulant expansion gives that $U = \sum^{L}_{j=1} Z_j$ (see (\ref{eq-simu-U-add})), based on which the simulation of $U$ can be simplified to first simulate $Z_j$, $j=1,2,\cdots, L$. 
In the following, we demonstrate that only simple operations are required to simulate $Z_j$ by exploiting \textit{Theorem 1}.  

\textit{Theorem 1}: Let $N$ be a Poisson RV, denoted as $N\sim \text{Pois}(\lambda_j)$. Its probability mass function is 
\begin{equation}  
	\begin{split} 
		& f_{N} (k) = \exp (-\lambda_j) \lambda_j^k / k!  
		\label{eq-simu-poiss}  
	\end{split} 
\end{equation} 
where $k$ is a nonnegative integer representing the realization of $N$. 
Let $E_n$, $n=1,2,\cdots,N$, be independent and identically distributed (i.i.d.) exponential RVs with the rate parameter $a_j$. That is, its PDF and LT are 
\begin{equation}  
	\begin{split} 
		& f_{E_n} (x) = a_j \exp (-a_jx), \enspace  \mathcal{L}_{E_n} (s) = a_j / (s + a_j).
		\label{eq-simu-exp}  
	\end{split} 
\end{equation} 
We have 
\begin{equation}  
	\begin{split} 
		& Z = \sum_{n=1}^N E_n \overset{d}{=} Z_j
		\label{eq-simu-Poisson-Exp}  
	\end{split} 
\end{equation} 
where $\overset{d}{=}$ denotes equality in distribution. 

\textit{Proof}: see Appendix B. 

\textit{Theorem 1} indicates that $Z_j$, $j=1,2,\cdots, L$, can be derived from sums of $N$ i.i.d. exponential RVs, where $N$ is a poisson RV. This implies that $Z_j$, $j=1,2,\cdots, L$, follow the conditional gamma distribution of integer shape parameter (i.e., the Erlang distribution). 
Specifically, the conditional $Z_j$ follows the gamma distribution of shape parameter $k$ and scale parameter $1/a_j$, denoted by   
\begin{equation}  
	\begin{split} 
		& Z_j  \mid N=k  \sim \text{GM} (k, 1/a_j), \enspace \text{with } N\sim \text{Pois}(\lambda_j)  
		\label{eq-simu-Erlang}  
	\end{split} 
\end{equation} 
where $Z_j=0$ when $k=0$. 

Thus, the sample of $Z_j$, $j=1,2,\cdots, L$, can be independently generated via well-established techniques for Poisson RVs and gamma RVs. Then, adding theses independent samples yields the simulated sample of $U$, i.e., $u(m)$. 

The above simulation of the AR input sequence is the core, based on which we give the whole AR process simulation procedure below. 

\subsection{Whole Procedure of AR Process Simulation} 
The procedure of the AR process simulation based on the cumulant-expansion analytic continuation is summarized as follows:  
\begin{itemize} 
	\item {Step 1: Establish an AR model based on the prescribed ACF/PSD using Yule-Walker method}; 
	\item {Step 2: Compute the unit impulse response $h_i$ in (\ref{eq-AR-ARMA-Green}) based on the coefficients of the AR model;} 
	\item {Step 3: Given $h_i$, compute the cumulants of the AR input $U$ via (\ref{eq-AR-LP-cum-rela}), and then construct $U$'s LT at zero via the cumulant expansion in (\ref{eq-AR-U-cum-logLP-series});} 
	\item {Step 4: Exploit the PA-based continuation of the cumulant expansion to obtain $U$'s LT over the entire complex plane via (\ref{eq-AC-Cum-LT}), defined by parameters $\lambda_j$ and $a_j$, $j=1,2,\cdots, L$;} 
	\item {Step 5: For each $j=1,2,\cdots, L$, independently generate Poisson sample $s_\text{pois}(m)$ following $N\sim \text{Pois}(\lambda_j)$; conditional on $N=s_\text{Pois}(m)$, generate gamma sample $z_{j}(m)$ following $Z_j\sim \text{GM} (s_\text{Pois}(m), 1/a_j)$;}  
	\item {Step 6: Add these samples yields the sample of $U$, i.e., $u(m)=\textstyle \sum_{j=1}^{L} z_{j}(m)$;}  
	\item {Step 7: Remove the mean from $u(m)$ and then input the zero-mean sample into the established AR model to obtain the output; add the desired mean back to the output to obtain $y(m)$.}   
\end{itemize} 

The above Steps 3-6 correspond to the simulation of the AR input sequence. 

Notably, the entire procedure does not involve any complicated operations, such as infinite series summation and numerical integration in ZMNL or the optimization required by maximum entropy-based approaches \cite{Wu2024-AR-ME,Zhao2025-AR-ME}.

\begin{figure}[t] 
	\centering
	\subfloat[]{ 
		\includegraphics[width=0.46\linewidth]{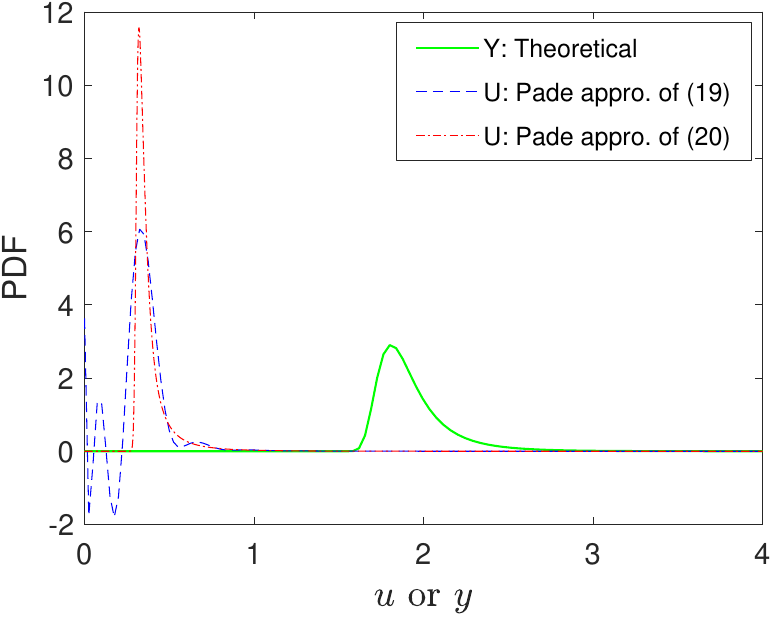} 
	} 
	\subfloat[]{ 
		\includegraphics[width=0.473\linewidth]{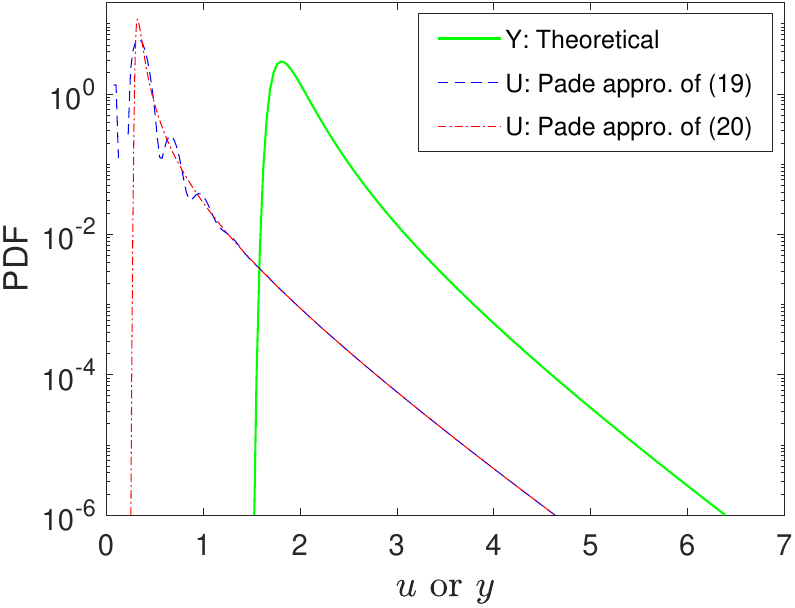}
	}	
	\caption{Comparison of the recovery performance of the AR input $U$'s PDF via the PA-based continuations of different series in Example 1. The AR output $Y$'s PDF is also given. (a) Linear scale; (b) Logarithmic scale.}  
	\label{fig-ARlow-diff-Pade-PDF} 
\end{figure} 
\begin{figure}[t] 
	\centering
	\subfloat[]{ 
		\includegraphics[width=0.465\linewidth]{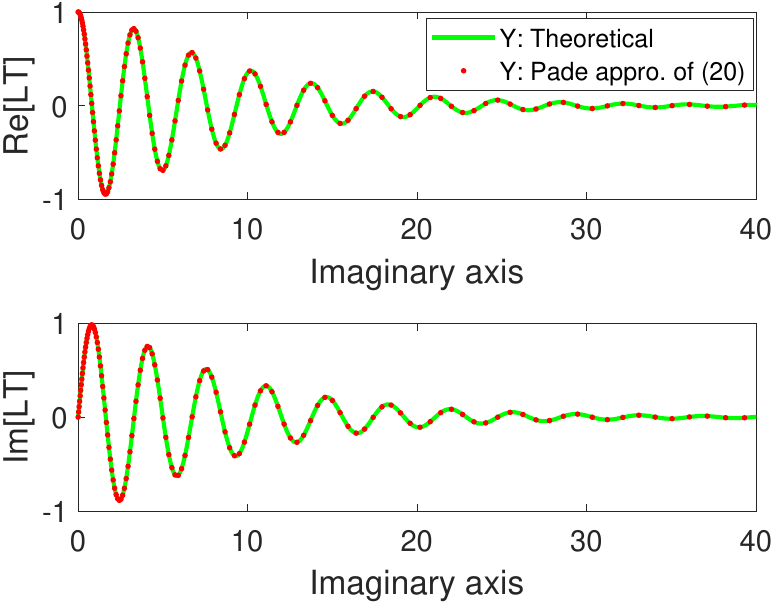} 
	} 
	\subfloat[]{ 
		\includegraphics[width=0.47\linewidth]{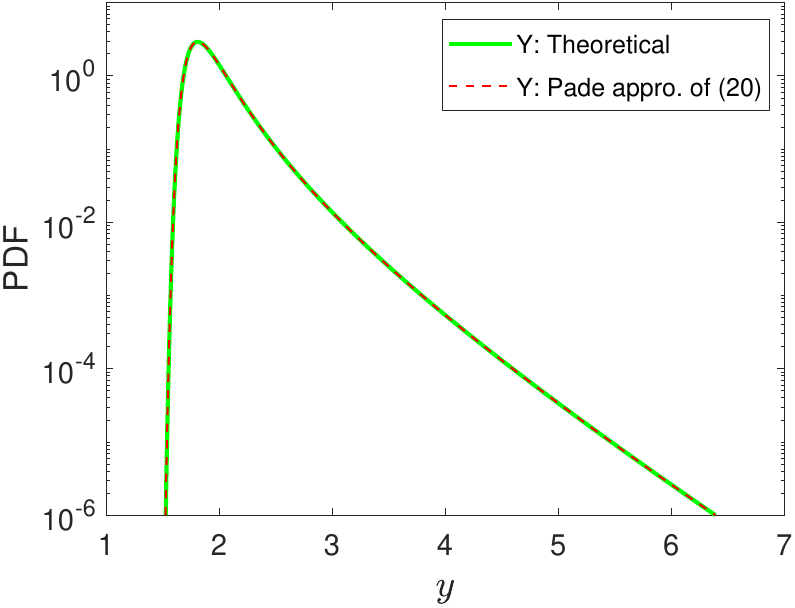}
	}	
	\caption{Comparison between the theoretical AR output $Y$ and that obtained by driving the AR process with $U$ generated via the PA-based continuation of the cumulant expansion. (a) LT; (b) PDF.}   
	\label{fig-ARlow-diff-Pade-PDF-veri-LT} 
\end{figure}
\begin{figure}[t] 
	\centering
	\subfloat[]{ 
		\includegraphics[width=0.472\linewidth]{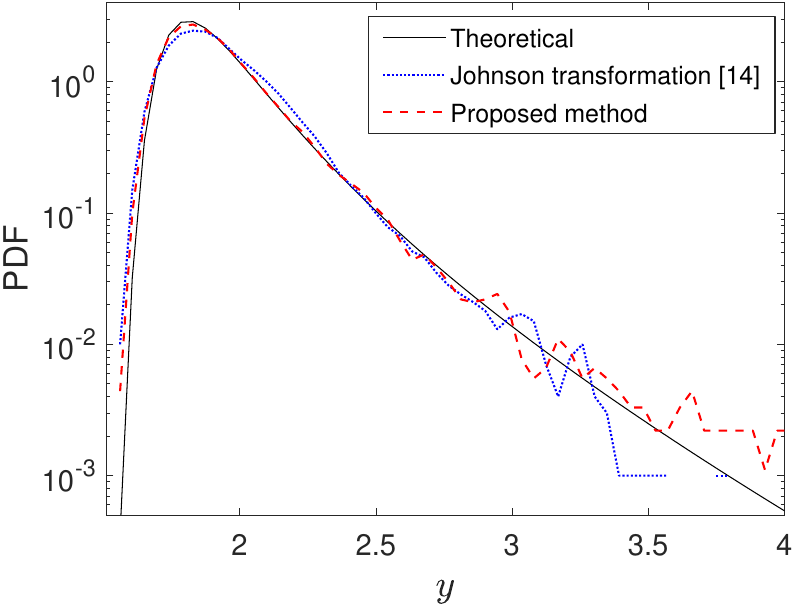} 
	} 
	\subfloat[]{ 
		\includegraphics[width=0.472\linewidth]{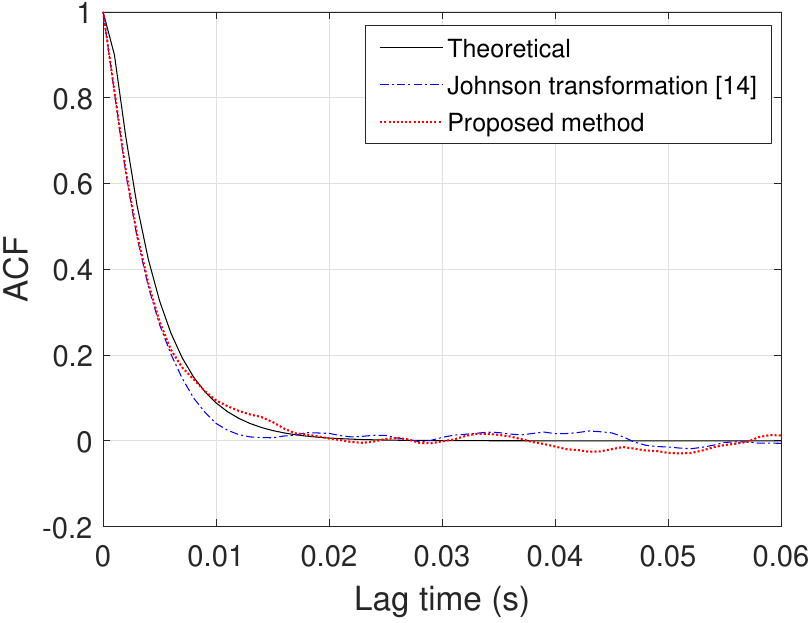}
	}	
	\caption{Simulated PT$\alpha$S texture in Example 1. (a) PDF; (b) ACF.}  
	\label{fig-ARlow-simu} 
\end{figure} 

\section{Performance Evaluation of Clutter Texture Simulation} 
\label{sec:Proposed-method-verify}   
To evaluate the performance of the AR process based on the cumulant-expansion analytic continuation in clutter texture simulation, two representative examples are considered, corresponding to textures with relatively light-tailed and heavy-tailed behaviors, respectively. In particular, the texture is modeled using the recently proposed PT$\alpha$S distribution \cite{Liao2024-PTaS}, which includes the commonly used gamma and inverse Gaussian clutter texture models as special cases.  

\subsection{PT$\alpha$S Texture with Relatively Light Tail}   
The first example corresponds to a relatively light-tailed texture, where the PT$\alpha$S variable $Y$ is specified with parameters $\alpha = 1.9$, $\gamma = 2$, and $\eta = 0.5$. Light-tailed textures are typically associated with relatively calm sea conditions, under which the texture exhibits weaker long-range dependence and less pronounced temporal variability. Consequently, a simple AR model is sufficient in this case, and an AR(2) model with parameters $a_1 = 0.9$ and $a_2 = -0.1$ is adopted. 
Other key simulation parameters are as follows: the pulse repetition frequency (PRF) is set to $F_a = 1000$Hz, and the length of the simulated texture sequence is $L_\text{sim} = 10000$

Fig. \ref{fig-ARlow-diff-Pade-PDF} compares the PDF recovery performance of the AR input $U$ obtained using the proposed PA-based continuation of the cumulant expansion. For comparison, the result based on the moment expansion is also included. 
As expected, the continuation of the cumulant expansion yields a valid PDF, whereas the continuation of the moment expansion results in an invalid PDF with negative values. 
This behavior can be attributed to their respective LT recovery performance, as analyzed in Section \ref{sec:Proposed-method}-C.

Using the AR input $U$ recovered via the PA-based cumulant-expansion continuation to drive the AR process, we can obtain the correponding LT and PDF of AR output $Y$, that is,   
\begin{equation}  
	\begin{split} 
		&{\mathcal{L}_{Y_\text{sim}}}( s ) = \sum\limits_{i = 1}^{L_\text{IR}} {\mathcal{L}_{U}}( h_i s )  = \sum\limits_{i = 1}^{L_\text{IR}} \prod_{j=1}^{L} \exp \left (  \frac{-\lambda_j h_i s}{h_i s+a_j} \right )   
		\label{eq-AR-output-Y-LT}  
	\end{split} 
\end{equation} 
\begin{equation}  
	\begin{split} 
		f_{Y_\text{sim}}(y) &= \mathcal{L}^{-1}\left[ \sum\limits_{i = 1}^{L_\text{IR}} \prod_{j=1}^{L} \exp \left (  \frac{-\lambda_j h_i s}{h_i s+a_j} \right )   \right](y)  
		\label{eq-AR-output-Y-PDF}  
	\end{split} 
\end{equation} 
where $y$ denotes the realization of $Y$.

Fig. \ref{fig-ARlow-diff-Pade-PDF-veri-LT} compares the resulting LT and PDF of $Y$ in (\ref{eq-AR-output-Y-LT}) and (\ref{eq-AR-output-Y-PDF}) with those of the prescribed PT$\alpha$S distribution, demonstrating excellent agreement in both the complex-valued LT and the PDF. 
These results confirm the effectiveness of the proposed PA-based continuation of the cumulant expansion.

To evaluate the actual simulation performance, Fig. \ref{fig-ARlow-simu} compares the PDF and ACF of the simulated PT$\alpha$S texture with theoretical values. 
For comparison evaluation, the efficient Johnson transformation method designed under the linear filtering framework \cite{Ma2019-new} is also adopted. 
The results show that the proposed method aligns better with the theoretical PDF in the body region compared to the Johnson transformation method, while both methods achieve similar ACF results.

\begin{table*}[t]
	\renewcommand\arraystretch{1.2} 
	\begin{center}
		\begin{threeparttable}  
			\caption{PDF and ACF MAEs of the simulated results relative to the theoretical values, along with the computational time of different methods, corresponding to Fig. \ref{fig-ARlow-simu} and Fig. \ref{fig-ARhigh-simu}} 
			\begin{tabular}{p{3.3cm}|p{1.4cm}<{\centering}p{1.4cm}<{\centering}p{1.1cm}<{\centering}|p{1.4cm}<{\centering}p{1.4cm}<{\centering}p{1.1cm}<{\centering}} 
				\toprule 
				\multirow{2}{*}{Methods}&\multicolumn{3}{c|}{Example 1} &\multicolumn{3}{c}{Example 2} \\ 
				\cline{2-4} 
				\cline{5-7} 
				&\multirow{1}{*}{PDF MAE} & \multirow{1}{*}{ACF MAE} & \multirow{1}{*}{Time (s)} & \multirow{1}{*}{PDF MAE} & \multirow{1}{*}{ACF MAE} & \multirow{1}{*}{Time (s)}\\ 
				\midrule
				Johnson transformation \cite{Ma2019-new}  &  0.0204 &  0.0182 &  0.0475 &  0.0534 & 0.0263 & 0.0436 \\   
				Proposed method      &  0.0067 &  0.0181 &  0.1155 & 0.0123 &  0.0263 &  0.0935 \\   
				\bottomrule  
			\end{tabular} 
			\begin{tablenotes}  
				\item[] \textit{Note:} Averages computed over 500 independent trials in each case. 
			\end{tablenotes} 
			\label{tabl-MonteCarlo} 
		\end{threeparttable} 
	\end{center} 
\end{table*} 
\begin{figure}[t] 
	\centering
	\subfloat[]{ 
		\includegraphics[width=0.457\linewidth]{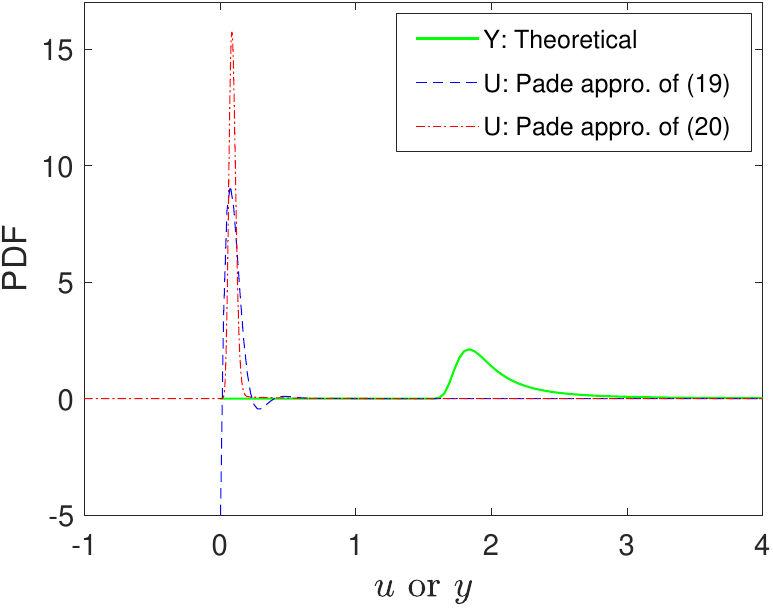} 
	} 
	\subfloat[]{ 
		\includegraphics[width=0.472\linewidth]{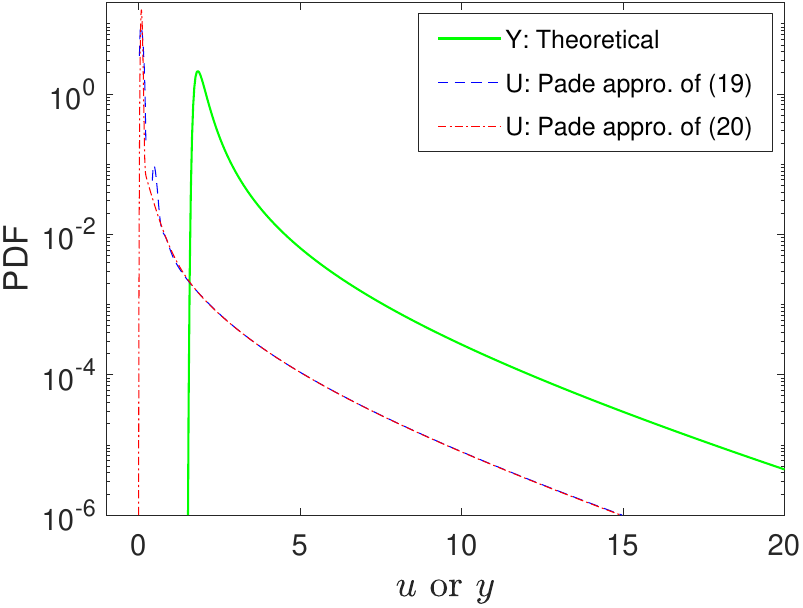}
	}	
	\caption{Comparison of the recovery performance of the AR input $U$'s PDF via the PA-based continuations of different series  in Example 2. The AR output $Y$'s PDF is also given. (a) Linear scale; (b) Logarithmic scale.}  
	\label{fig-ARhigh-diff-Pade-PDF} 
\end{figure} 
\begin{figure}[t] 
	\centering
	\subfloat[]{ 
		\includegraphics[width=0.46\linewidth]{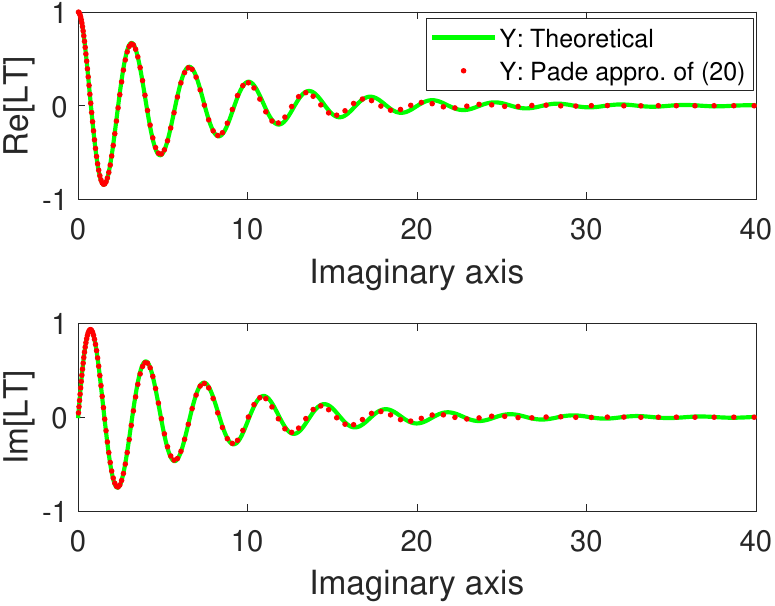} 
	} 
	\subfloat[]{ 
		\includegraphics[width=0.472\linewidth]{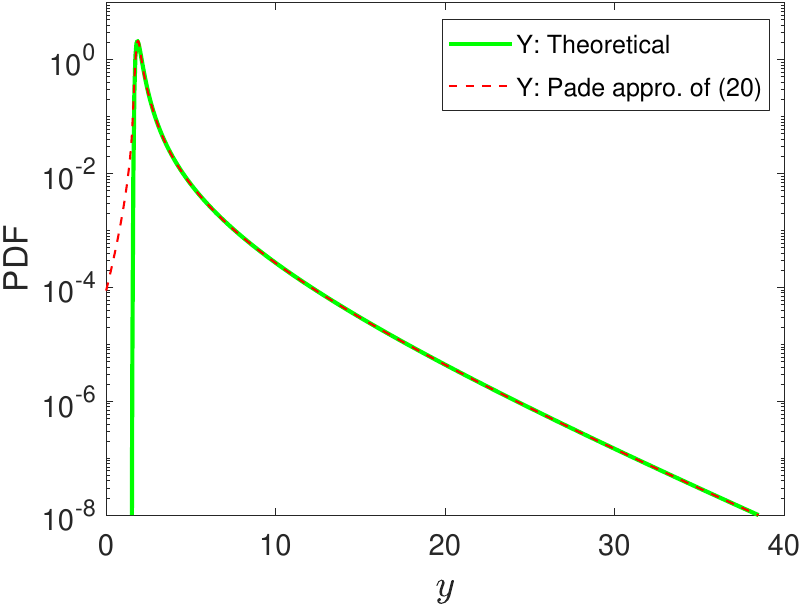}
	}	
	\caption{Comparison between the theoretical AR output $Y$ and that obtained by driving the AR process with $U$ generated via the PA-based continuation of the cumulant expansion in Example 2. (a) LT; (b) PDF.}  
	\label{fig-ARhigh-diff-Pade-PDF-veri-LT} 
\end{figure} 
\begin{figure}[t] 
	\centering 
	\subfloat[]{ 
		\includegraphics[width=0.476\linewidth]{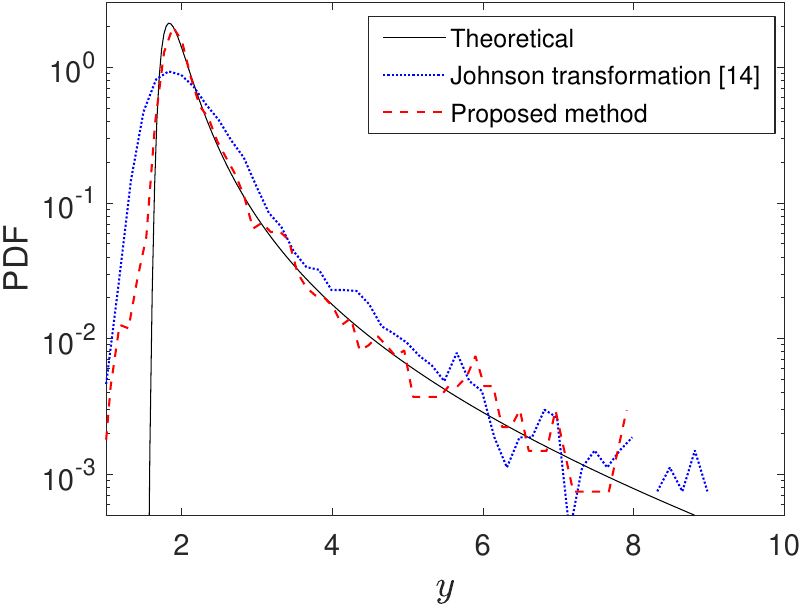} 
	} 
	\subfloat[]{ 
		\includegraphics[width=0.465\linewidth]{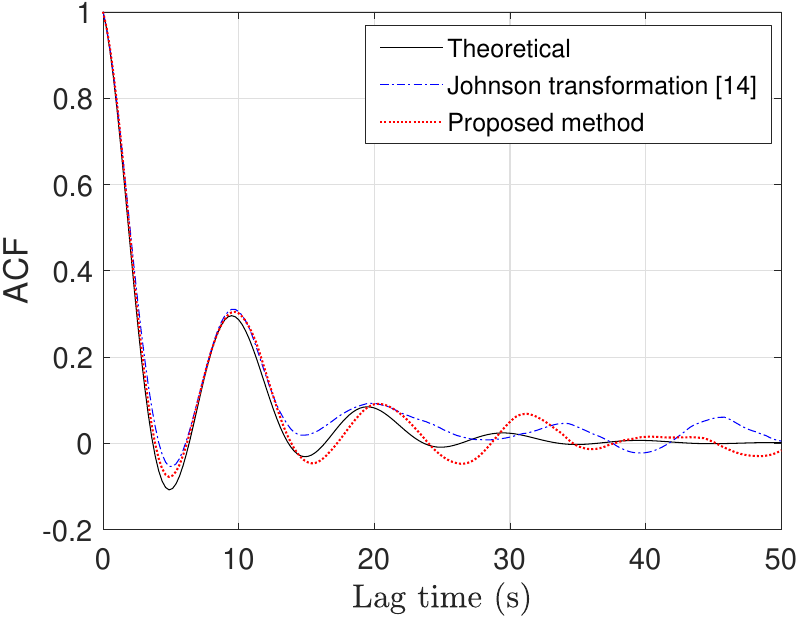}
	}	
	\caption{Simulated PT$\alpha$S texture in Example 2. (a) PDF; (b) ACF.}  
	\label{fig-ARhigh-simu} 
\end{figure} 

\subsection{PT$\alpha$S Texture with Relatively Heavy Tail}    
The second example corresponds to a relatively heavy-tailed texture, where the PT$\alpha$S variable $Y$ is specified with parameters $\alpha = 1.9$, $\gamma = 2$, and $\eta = 4$. 
In heavy-tailed case, there exists long-range dependence and pronounced temporal variability, and the texture temporal correlation can be modeled using an exponential attenuation function modulated by a cosine term \cite{Wang2015-AP}  
\begin{equation}  
	\begin{split} 
		& R_\text{tex}(\tau)=\exp \left(-\frac{\tau}{t_{0}}\right)\left[d_\text{mod} \cos \left(\frac{2 \pi \tau}{T_{0}}\right)+(1-d_\text{mod})\right] 
		\label{eq-corr-texture} 
	\end{split} 
\end{equation} 
where $t_{0}$ is the texture decorrelation time, $T_{0}$ is the variation period, and $d_\text{mod}$ ($0 < d_\text{mod} < 1$) is a constant representing the influence of the cosine function.  
To adequately reflect fluctuation effects, the texture modulation parameter is set to $d_{\mathrm{mod}} = 0.6$, decorrelation time $t_{0}=8$s, and variation period $T_{0}=10$s. Accordingly, a higher-order AR model is required to represent the modulation introduced by the cosine term. 
Other key simulation parameters are as follows: $F_a = 1000$Hz and $L_\text{sim} = 10000$.

Fig. \ref{fig-ARhigh-diff-Pade-PDF} compares the PDF recovery performance of the AR input $U$ obtained using the PA-based continuations of different series. Similar to Example 1, only the proposed continuation based on the cumulant expansion yields a valid PDF. 
Further, Fig. \ref{fig-ARhigh-diff-Pade-PDF-veri-LT} compares the LT and PDF of the corresponding AR output $Y$ with the prescribed ones. It shows excellent agreement in both the complex-valued LT and the PDF, indicating the effectiveness of the proposed PA-based continuation of cumulant expansion.

For actual simulation evaluation, Fig. \ref{fig-ARhigh-simu} compares the PDF and ACF of the simulated PT$\alpha$S texture with their theoretical counterparts for Example 2.
The results indicate that the proposed method matches the prescribed PDF in the body and tail regions much more accurately than the Johnson transformation method, while both methods achieve comparable accuracy in reproducing the given ACF. 
In particular, when comparing the simulation performance of the two methods under different tail behaviors, it is observed that the Johnson transformation method provides satisfactory performance for relatively light-tailed clutter (see Fig. \ref{fig-ARlow-simu}), but degrades significantly for heavy-tailed clutter (see Fig. \ref{fig-ARhigh-simu}), whereas the proposed method consistently delivers satisfactory performance.  
This difference can be attributed to the fact that the Johnson transformation method is limited to using only the first four moments, whereas the proposed method can easily exploit a much richer set of cumulants (moments).

The observations in Figs. \ref{fig-ARlow-simu} and \ref{fig-ARhigh-simu} are further corroborated by Table \ref{tabl-MonteCarlo}, which reports the average mean absolute errors (MAEs) of the PDF and ACF over 500 independent Monte Carlo trials for both examples. 
The MAE is calculated using
\begin{equation} 
\begin{split} 
&MAE  = \frac{1}{L_\text{err}}\sum_{i=1}^{L_\text{err}}   \left|  D_{\text{theo},i}  -  D_{\text{theo},i}   \right|.   
\label{eq-MSE} 
\end{split} 
\end{equation} 
For computing PDF MAE, $L_\text{err}=\sqrt{L_\text{sim}}$ denotes the number of histogram bins, with $D_{\text{theo},i}$ and $D_{\text{dat},i}$ representing the theoretical and simulated PDF value for each bin. 
For computing ACF MAE, $L_\text{err}=200$ denotes the lag number, with $D_{\text{theo},i}$ and $D_{\text{dat},i}$ denoting the theoretical and simulated ACF values at each lag.

Moreover, the average consuming times of both methods are recorded. 
As shown, both methods are highly efficient in simulating a sequence of length $L_\text{sim} = 10000$. In particular, the Johnson transformation method, which involves only very simple operations, is more than twice as fast as the proposed method.
Nevertheless, it is noteworthy that the proposed method also relies on relatively simple operations, with the most computationally demanding step being the inversion of a matrix of order less than 20, which can be efficiently handled on various platforms.
The moderately longer computational time of the proposed method is mainly due to the cyclic generation of $z_j(m)$, $j=1,2,\cdots,L$ (see Section \ref{sec:Proposed-method-simu}-B, Step 5). 
However, this step is amenable to parallel implementation, making further acceleration straightforward.

\section{Conclusion}  
\label{sec:conclusion}   
This paper proposes a series-based analytic continuation strategy to enhance AR process simulation for non-Gaussian processes. 
Specifically, the PA-based continuation of the cumulant expansion in the logarithmic LT domain enables accurate recovery of the AR input's statistics, and supports fast simulation relying on only simple operations.  
Compared with existing AR simulation methods, the proposed method more effectively exploits moment information to deliver accurate simulation with simple operations; and relative to the ZMNL method, it is particularly suitable for models lacking a closed-form inverse CDF. 
Simulation results demonstrate that the proposed method reliably reproduces the prescribed PDF and ACF for PT$\alpha$S clutter textures across both light-tailed and heavy-tailed cases. By comparison, the Johnson transformation method, although relatively faster, shows significant deviation from the theoretical PDF when applied to heavy-tailed PT$\alpha$S clutter textures. 

In this study, we focus on simulating one-dimensional real-valued sequence with prescribed PDF and ACF. 
Notably, this serves as the foundation for naturally extensions to complex-valued, non-stationary, or two-dimensional clutter simulations, as outlined below: 
\begin{itemize} 
\item 
Although only the simulated clutter texture is presented in the performance analysis section, the corresponding complex-valued CG clutter can be readily obtained by multiplying the texture with complex Gaussian speckle. 

\item 
The proposed method focuses on stationary clutter; however, it can be naturally extended to spatially or temporally non-stationary clutter by incorporating an evolving Doppler spectrum model \cite{Watts2012-EDSM}. 

\item 
Owing to the additive and scaling properties of cumulants under linear filtering, the proposed method admits a straightforward extension to the multivariable AR process. A brief analysis is presented in Appendix C. 
\end{itemize} 

Moreover, the proposed method via series-based analytic continuation also applies to other applications, such as financial analysis and structural engineering, for the simulation of non-Gaussian correlated sequences. 
However, it should be noted that it requires moments of the AR output to be finite, and thus need further modification for distributions with divergent moments, such as the inverse Gamma distribution.

\appendices 
\section{Basics of the Pad\'{e} Approximation (PA)}
Let $h$ be analytic in a neighborhood of $z = 0$. Then, $h(z)$ can be expressed as a Taylor series in the complex variable $z$ given by 
\begin{equation}  
	\begin{split} 
		h( z ) &= \sum\limits_{n = 0}^\infty    {c_{n}}  {{z}^n}    
		\label{eq-A1} 
	\end{split} 
\end{equation} 

This series defines the function within its region of convergence and serves as the basis for analytic continuation beyond that region.
The PA achieves such continuation by constructing a rational approximation to $h(z)$ with numerator degree $K$ and denominator degree $L$, whose Taylor expansion about $z=0$ matches that of $h(z)$ up to order $L+M$. 
Specifically, the PA of order $[K,L]$ for the series $h(z)$ is given by the rational function as 
\begin{equation}  
	\begin{split} 
		{[K,L]}_h( z ) & = {\sum\limits_{i = 0}^K  P_i z^i}/{\sum\limits_{j = 0}^L  Q_j z^j}   
		\label{eq-A2} 
	\end{split} 
\end{equation} 
where the coefficients ${P_i}$ and ${Q_j}$ are determined such that 
\begin{equation}  
	\begin{split} 
		\biggl( \sum_{n=0}^{K+L} c_n z^n \biggr) 
		\biggl( \sum_{j=0}^{L} Q_j z^j \biggr) 
		- \sum_{i=0}^{K} P_i z^i 
		= \mathcal{O}\bigl( z^{K+L+1} \bigr) 
		\label{eq-A3} 
	\end{split} 
\end{equation} 
where $\mathcal{O}\bigl( z^{K+L+1} \bigr)$ accounts for terms of order higher than $z^{K+L}$. 
In particular, we fix the coefficient $Q_0=1$ without loss of generality. 

Eq. (\ref{eq-A3}) corresponds to the moment-matching approach. 
By cross-multiplying, it can be written explicitly in terms of the coefficients ${Q_j}$and ${P_i}$.
In particular, the denominator coefficients are related to the power-series coefficients through 
\begin{equation}  
	\begin{split} 
		\sum_{j=1}^{L} c_{K+i-j}  Q_j = c_{K+i}, \enspace 1 \leq i,j \leq L   
		\label{eq-A4} 
	\end{split} 
\end{equation}
which form a set of $L$ linear equations for the $L$ unknown denominator coefficients. Thus, the denominator coefficients are obtained by forming and solving the Hankel matrix and linear system as 
\begin{equation}  
	\small
	\begin{split} 
		&{\left[\begin{array}{cccc}
				c_{K-L+1} & c_{K-L+2} & \cdots & c_{K} \\
				\vdots & \vdots & \vdots & \vdots \\
				c_{K-L+i} & c_{K-L+i+1} & \cdots & c_{K+i-1} \\
				\vdots & \vdots & \vdots & \vdots \\
				c_{K} & c_{K+1} & \cdots & c_{K+L-1}
			\end{array}\right]} {\left[\begin{array}{c}
				Q_{L}  \\
				\vdots  \\
				Q_{j}  \\
				\vdots \\
				Q_{1}
			\end{array}\right]} \\
		& =  {\left[\begin{array}{c}
				- c_{K+1}  \\
				\vdots  \\
				-c_{K+i}  \\
				\vdots \\
				-c_{K+L}
			\end{array}\right]}. 
		\label{eq-A4-1}  
	\end{split} 
\end{equation}

The coefficients in the numerator are instead found from (\ref{eq-A3}) as follows: 
\begin{equation}  
	\begin{split} 
		P_n =  \sum_{j=0}^{n}  c_{n-j} Q_j, \enspace 0 \leq j\leq K 
		\label{eq-A5} 
	\end{split} 
\end{equation} 
where $Q_j = 0$ if $j\ge L$.  

In summary, (\ref{eq-A4}) and (\ref{eq-A5}) determine the PA for a function $h(z)$ given its power series expansion (\ref{eq-A1}). 
Practical implementations of the PA are available in most standard mathematical computing environments, including Mathematica, MATLAB, and Maple. 

\section{Proof of \textit{Theorem 1}}
Based on the definition of $Z$ in (\ref{eq-simu-Poisson-Exp}), we can get its conditional LT as
\begin{equation}  
	\begin{split} 
		& \mathbb{E}\left[e^{-s Z} \mid N=k\right]=\left(\frac{a_j}{s+a_j}\right)^{k}   
		\label{eq-simu-proof1}  
	\end{split} 
\end{equation} 
where $\mathbb{E}\left[ \cdot \right]$ denotes the expectation operator. 

Then, the (unconditional) LT can be derived via
\begin{equation}  
	\begin{split} 
		\mathcal{L}_{Z}(s) = \mathbb{E}\left[e^{-s Z}\right] & =\sum_{k=0}^{\infty} \mathbb{P}(N=k) \mathbb{E}\left[e^{-s Z} \mid N=k\right] \\
		& =\sum_{k=0}^{\infty} \exp (-\lambda_j) \frac{\lambda_j^{k}}{k!}\left(\frac{a_j}{s+a_j}\right)^{k} \\
		& = \exp (-\lambda_j) \sum_{k=0}^{\infty} \left(\frac{a_j\lambda_j}{s+a_j}\right)^{k} / {k!}
		\label{eq-simu-proof2}  
	\end{split} 
\end{equation} 
where $\mathbb{P}\left( \cdot \right) $ denotes the probability of an event. 
Since the series in (\ref{eq-simu-proof2}) is the generating function of the exponential distribution, we can have
\begin{equation}  
	\begin{split} 
		\mathcal{L}_{Z}(s) &=  \exp  \left ( -\lambda_j  \right ) \exp  \left (  \frac{a_j \lambda_j}{s + a_j}  \right )  =  \exp  \left ( \frac{-\lambda_j s}{s + a_j}  \right ).  
		\label{eq-simu-proof3}  
	\end{split} 
\end{equation} 
The result is the same as the LT of $Z_j$ given in (\ref{eq-simu-U-add}), which completes the proof.

\section{The Proposed Series-Based Analytic Continuation for Multivariate AR Process Simulation} 
For a multivariate AR or ARMA process, let the output vector be defined as $\mathbf{Y}(m)=\left [ y_1(m),y_2(m),\cdots, y_M(m)\right ]$, which comprises $M$ mutually correlated univariate non-Gaussian correlated processes. 
Similarly, define the input vector as $\mathbf{U}(m)=\left [u_1(m),u_2(m),\cdots, u_M(m) \right] $, which consists of $M$ mutually independent univariate non-Gaussian white processes. 
Then, the multivariate AR or ARMA process can be rewritten in a form of \cite{Ma2019-new} 
\begin{equation}  
	\begin{split} 
		{y_p}\left( m \right)  &= \sum\limits_{i = 0}^{\infty} {\sum\limits_{q = 1}^M {{h_{i,p,q}}{u_q}\left( {m - i} \right)} }\\
		&\simeq \sum\limits_{i = 0}^{L_\text{IR}} {\sum\limits_{q = 1}^M {{h_{i,p,q}}{u_q}\left( {m - i} \right)} }, \enspace p = 1,2,\cdots M 
		\label{eq-mul-AR-1}  
	\end{split} 
\end{equation} 
where $h_{i,p,q}$ denotes the $(i,p,q)$-th unit impulse response of an $L_\text{IR}\times M \times M$ impulse-response tensor computed from the prescribed ACF and cross-correlation function of the output vector $\mathbf{Y}$.  

Since $u_q$, $q=1,2,\cdots,M$, are mutually independent, and $u_q(m)$ are i.i.d. over time,  the LT of $y_p$ satisfies  
\begin{equation}  
	\begin{split} 
		{{\cal L}_{{y_p}}}\left( s \right) = \prod\limits_{i = 0}^{L_\text{IR}} {\prod\limits_{q = 1}^M {{{\cal L}_{{u_q}}}\left( {{h_{i,p,q}}s} \right)} } , \enspace p = 1,2,\cdots M. 
		\label{eq-mul-AR-2}  
	\end{split} 
\end{equation} 

Based on (\ref{eq-mul-AR-2}), the $n$-th cumulants of $y_p$ can be expressed as  
\begin{equation}  
	\begin{split} 
		{\kappa _{{y_p},n}} &= {\left( { - 1} \right)^n}\frac{{{d^n}}}{{d{\xi ^n}}}\ln {{\cal L}_{{y_r}}}\left( s \right){|_{s = 0}} \\
		&=\sum\limits_{i = 0}^{L_\text{IR}}  {\sum\limits_{p = 1}^M {{{\left( { - 1} \right)}^n}\frac{{{d^n}}}{{d{s^n}}}\ln {{\cal L}_{{u_p}}}({h_{i,p,q}}s)} } {|_{s = 0}} \\
		& = \sum\limits_{i = 0}^{L_\text{IR}}  {\sum\limits_{p = 1}^M {h_{i,p,q}^n{\kappa _{{u_p},n}}} } , \enspace p = 1,2,\cdots M.  
		\label{eq-mul-AR-3}  
	\end{split} 
\end{equation} 

Thus, the $n$-th cumulants of $u_q$, $q = 1,2,\cdots M$ can be computed from a linear system of equations as
\begin{equation}  
	\begin{split} 
		&\begin{bmatrix}
			\kappa _{{u_1},n} \\
			\kappa _{{u_2},n}\\
			\vdots \\
			\kappa _{{u_M},n}
		\end{bmatrix}=\\
		&\begin{bmatrix}
			\sum\limits_{i = 0}^{L_\text{IR}}  {h_{i,1,1}^n} &  \sum\limits_{i = 0}^{L_\text{IR}}  {h_{i,1,2}^n}&  \cdots& \sum\limits_{i = 0}^{L_\text{IR}}  {h_{i,1,M}^n} \\
			\sum\limits_{i = 0}^{L_\text{IR}}  {h_{i,2,1}^n} &  \sum\limits_{i = 0}^{L_\text{IR}}  {h_{i,2,2}^n}&  \cdots& \sum\limits_{i = 0}^{L_\text{IR}}  {h_{i,2,M}^n} \\
			\vdots & \vdots&  \ddots &\vdots \\
			\sum\limits_{i = 0}^{L_\text{IR}}  {h_{i,M,1}^n} &  \sum\limits_{i = 0}^{L_\text{IR}}  {h_{i,M,2}^n}&  \cdots& \sum\limits_{i = 0}^{L_\text{IR}}  {h_{i,M,M}^n} \\
		\end{bmatrix}^{-1} 
		\begin{bmatrix}
		\kappa _{{y_1},n} \\
		\kappa _{{y_2},n}\\
		\vdots \\
		\kappa _{{y_M},n}
		\end{bmatrix}. 
		\label{eq-mul-AR-4}  
	\end{split} 
\end{equation} 

Subsequently, each sequence $u_q(m)$, $q = 1,2,\cdots M$, can be generated via the proposed series-based analytic continuation strategy, following the simulation procedure described in Section \ref{sec:Proposed-method-simu}-B. 
Finally, the simulation of $\mathbf{Y}(m)$ is completed by passing $u_q(m)$, $q = 1,2,\cdots M$, through the multivariate AR process in (\ref{eq-mul-AR-1}).

\section*{Acknowledgment}  
The authors would like to thank the Associate Editor and the reviewers for the valuable comments and suggestions to improve the quality of this article.


\end{document}